\documentclass[compsoc, conference, a4paper, 10pt, times]{IEEEtran}

\usepackage{cite}
\usepackage{amsmath,amssymb,amsfonts}
\usepackage{graphicx}
\usepackage{textcomp}
\usepackage{xcolor}
\usepackage{booktabs}
\usepackage{subcaption}
\usepackage{float}
\usepackage[hidelinks]{hyperref}
\usepackage{amsmath}
\usepackage{mdframed}
\usepackage{amsmath}
\usepackage{soul}

\begin{document}

\title{PASA: Attack Agnostic Unsupervised Adversarial Detection using Prediction \& Attribution Sensitivity Analysis}


\author{\IEEEauthorblockN{Dipkamal Bhusal}
\IEEEauthorblockA{\textit{Rochester Institute of Technology} \\
Rochester, NY, USA }
\and
\IEEEauthorblockN{Md Tanvirul Alam}
\IEEEauthorblockA{\textit{Rochester Institute of Technology} \\
Rochester, NY, USA }
\and
\IEEEauthorblockN{Monish K. Veerabhadran}
\IEEEauthorblockA{\textit{Rochester Institute of Technology} \\
Rochester, NY, USA }
\and
\IEEEauthorblockN{\hspace{10mm}Michael Clifford}
\IEEEauthorblockA{\hspace{10mm}\textit{Toyota Motor North America} \\
\hspace{10mm}Mountain View, CA, USA}
\and
\IEEEauthorblockN{\hspace{13mm}Sara Rampazzi}
\IEEEauthorblockA{\hspace{13mm}\textit{University of Florida} \\
\hspace{13mm}Gainesville, FL, USA }
\and
\IEEEauthorblockN{\hspace{13mm}Nidhi Rastogi}
\IEEEauthorblockA{\hspace{13mm}\textit{Rochester Institute of Technology} \\
\hspace{13mm}Rochester, NY, USA}}

\maketitle

\thispagestyle{plain}
\pagestyle{plain}

\begin{abstract}
Deep neural networks for classification are vulnerable to adversarial attacks, where small perturbations to input samples lead to incorrect predictions. This susceptibility, combined with the black-box nature of such networks, limits their adoption in critical applications like autonomous driving. Feature-attribution-based explanation methods provide relevance of input features for model predictions on input samples, thus explaining model decisions. However, we observe that both model predictions and feature attributions for input samples are sensitive to noise. We develop a practical method for this characteristic of model prediction and feature attribution to detect adversarial samples. Our method, PASA, requires the computation of two test statistics using model prediction and feature attribution and can reliably detect adversarial samples using thresholds learned from benign samples. We validate our lightweight approach by evaluating the performance of PASA on varying strengths of FGSM, PGD, BIM, and CW attacks on multiple image and non-image datasets. On average, we outperform state-of-the-art statistical unsupervised adversarial detectors on CIFAR-10 and ImageNet by 14\% and 35\% ROC-AUC scores, respectively. Moreover, our approach demonstrates competitive performance even when an adversary is aware of the defense mechanism.
\end{abstract}

\section{Introduction}
 Deep neural networks (DNNs) have demonstrated state-of-the-art performance in various classification tasks \cite{he2015delving, bojarski2016end, anderson2018ember}. However, DNNs are known to be vulnerable to adversarial evasion attacks. Attackers carefully and deliberately craft samples by adding small perturbations to fool the DNN and cause it to make incorrect predictions \cite{goodfellow2015explaining,  carlini2017towards, madry2017towards}. The susceptibility of DNNs to such attacks poses serious risks when deploying them in application scenarios where security and reliability are essential, such as in autonomous vehicles \cite{kumar2020black} and medical diagnosis\cite{shen2017deep}. 

Current approaches for defending against such evasion attacks can be broken into two broad categories. One category increases the robustness of a model (e.g., adversarial training \cite{goodfellow2015explaining}, feature denoising \cite{xie2019feature}). However, such approaches achieve model robustness at the cost of modification of the network architecture or training process and compromise natural accuracy as a result. Such methods are still susceptible to adversarial attacks like blind-spot attacks \cite{zhang2018limitations}. The other category identifies adversarial samples instead of making robust classifications. While detecting adversarial attacks is as challenging as classifying them \cite{tramer2022detecting}, such methods are useful in many practical situations where discarding adversarial samples for security or generating an alert for human intervention is possible. Detection methods are \textit{supervised} if they require both benign and adversarial samples in their training \cite{feinman2017detecting, ma2018characterizing}. The main limitations of supervised detection methods are the requirement of prior attack knowledge and the availability of adversarial samples. In contrast, \textit{unsupervised} methods solely rely on properties of natural (benign) samples for training \cite{meng2017magnet,xu2017feature}.

\begin{figure}[h!]
\centering{
     \resizebox{0.92\columnwidth}{!}{
\includegraphics{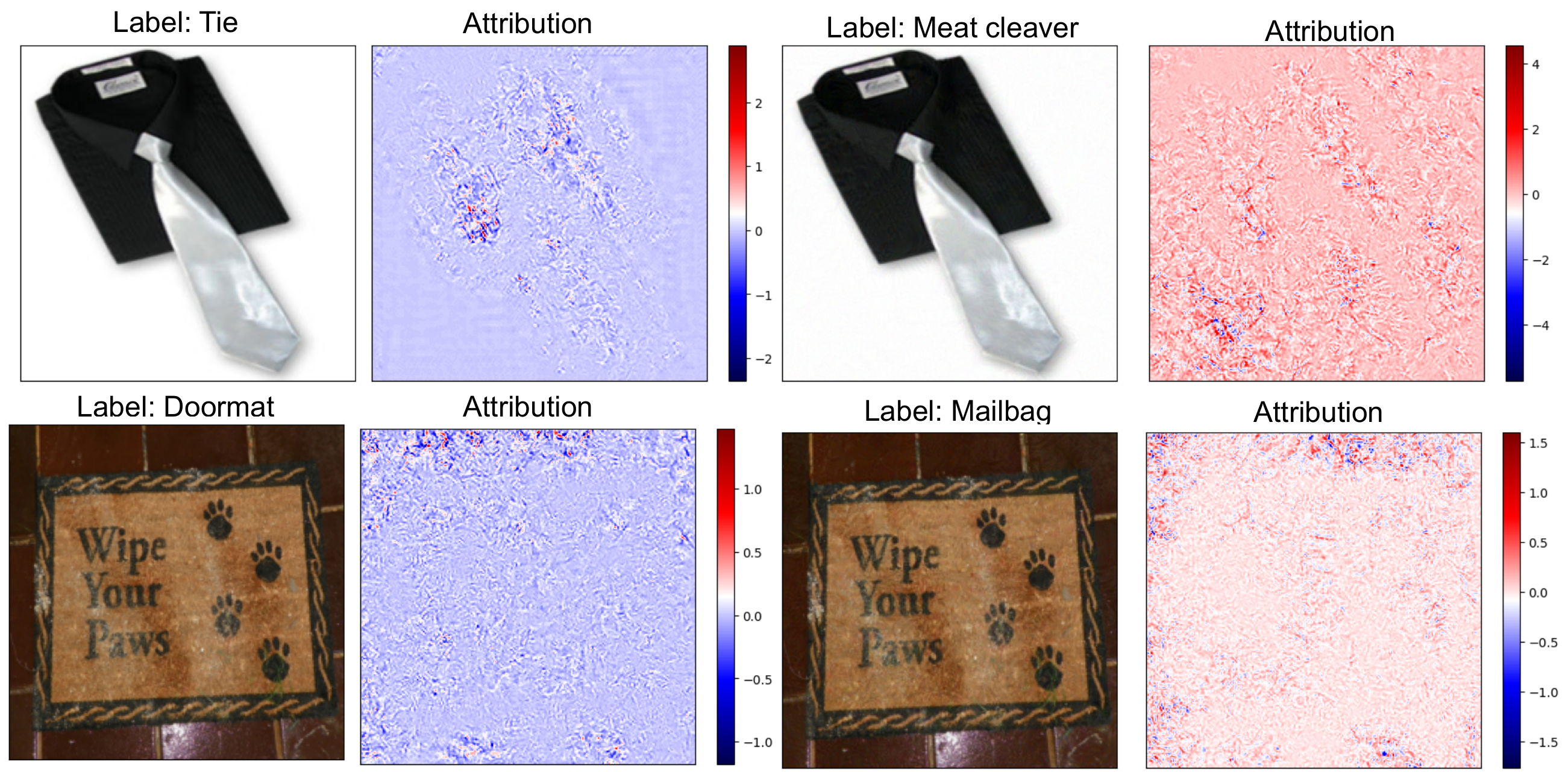}}
\caption[Caption]{Benign (1st column) and Adversarial PGD Image (3rd column). Corresponding Integrated Gradient (IG) Attribution (2nd and 4th column).}
    \label{fig:advSample}
    }
\end{figure}

A different line of research, called post-hoc explanation methods, addresses the black-box nature of DNNs \cite{simonyan2013deep, ribeiro2016should,sundararajan2017axiomatic, lundberg2017unified, bykov2022noisegrad}. Post-hoc explanation methods \textit{explain} the decision made by a DNN for a test input based on its input features, and enhance our understanding of the DNN's decision-making process. For example, in an image classifier, such methods can identify the key pixels in an input image that lead to a DNN decision. Explanations can use various approaches such as feature attribution, rules, or counterfactuals to explain an instance \cite{Bhusal2022SoKME}. \textit{Feature attribution-based methods}, such as  Integrated Gradient (IG) \cite{sundararajan2017axiomatic}, assign attribution or relevance scores to each input feature, quantifying the importance of the feature to the model's prediction. Recent research has explored the application of feature-attribution-based explanation methods in detecting adversarial attacks \cite{tao2018attacks, zhang2018detecting, yang2020ml,wang2020interpretability, vardhan2021exad}. However, these methods require both benign and adversarial samples to train an \textit{additional} classifier for detection and do not incorporate features from the classification model in the detection pipeline.

\textbf{Proposed Approach:} We propose a novel method for detecting adversarial samples by combining model prediction and feature attribution. Our approach is motivated by the evident differences in model prediction and feature attribution between benign and adversarial samples when noise is introduced: (a) we observe that a DNN exhibits distinct behavior when noise (e.g., gaussian noise) is introduced to adversarial samples, similar to studies performed by Roth et al. \cite{roth2019odds} and Hu et al. \cite{hu2019new}. However, while prior works \cite{roth2019odds, hu2019new} look for noise that does not change the model prediction for benign samples, we empirically identify noise that maximizes the distinction between the behavior of benign and adversarial samples. (b) There are \textit{noticeable differences} in the attribution map of benign and adversarial samples. Figure \ref{fig:advSample} illustrates these differences in images where the distribution of attribution scores varies significantly for adversarial samples, evident from the change in red and blue pixels in the attribution map. Even though adversarial samples are crafted by adding small perturbations to the input data, we observe that their feature attribution differs markedly from those of benign samples. This distinction in feature attribution becomes more pronounced when noise is introduced to samples. (c) Examining these discrepancies in model prediction and feature attribution of benign and adversarial samples subjected to additional perturbation can effectively detect adversarial attacks, and ensure the security of systems incorporating deep learning models.

We introduce PASA\footnote{``PASA" is the Newari term for ``friend," a Sino-Tibetan language spoken by the indigenous people of the Kathmandu Valley, Nepal.}, a threshold-based, unsupervised method for detecting adversarial samples, using \underline{\textbf{P}}rediction \& \underline{\textbf{A}}ttribution \underline{\textbf{S}}ensitivity \underline{\textbf{A}}nalysis. We use noise as a probe to modify input samples, measure changes in model prediction and feature attribution, and learn thresholds from benign samples. At test time, PASA computes model prediction and feature attribution of a given input and its noisy counterpart and rejects the input if the change in model prediction or feature attribution does not meet the predefined threshold. {We demonstrate the effectiveness of our lightweight detection approach by evaluating its performance on five different datasets (MNIST \cite{lecun1998gradient}, CIFAR-10 \cite{krizhevsky2009learning}, CIFAR-100 \cite{krizhevsky2009learning}, ImageNet \cite{deng2009imagenet} and updated CIC-IDS2017 \cite{engelen2021troubleshooting}) and five different deep neural network architectures (MLP \cite{rumelhart1985llearning}, LeNet \cite{lecun1998gradient}, VGG16 \cite{krizhevsky2009learning}, ResNet \cite{he2016deep}, and MobileNet \cite{sandler2018mobilenetv2}).} On average, PASA outperforms other state-of-the-art statistical unsupervised detectors (FS \cite{xu2017feature}, MagNet \cite{meng2017magnet}, LOO \cite{yang2020ml}, TWS \cite{hu2019new}) by 14\% on CIFAR-10, 4\% on CIFAR-100 and 35\% on ImageNet. We further evaluate PASA under an adaptive adversary setting and demonstrate its robustness. We observe that performing adaptive attacks against both model prediction and feature attribution increases computational complexity for an adversary, and PASA still achieves competitive performance. PASA has low inference latency, and the simplicity yet effectiveness of this approach makes it suitable for deployment in scenarios with limited computational resources. Our code is available at \textit{\url{{https://github.com/dipkamal/PASA}}}. %


\section{Background and Related Work}\label{sec:background}

\textbf{Deep Neural Network:} Deep neural networks (DNNs) learn efficient representations of training data by extracting features using interconnected neurons. Let $(X,Y)$ represent the training data where $X$ reflects input space and $Y$ reflects label space. Then, a deep neural network ($F$) generates a highly accurate functional representation $F:X\rightarrow Y$. Such networks are trained using backpropagation, a gradient-based optimization algorithm, which adjusts the weights between neurons to minimize the error between model predictions and ground truth labels. For example: Convolutional Neural Networks (CNNs) are a type of DNN used for image, video, tabular, and text data \cite{lecun1998gradient}. In supervised classification, a CNN is trained using $N$ labeled samples. The $i^{th}$ sample $({\textbf{x}_i}, y_i)$ consists of an input $\textbf{x}_i$ with label $y_i$. The final layer of the network is often referred to as a ``logit'' layer and consists of $k$ neurons corresponding to the $k$ target classes. The ``logits,'' $Z(\textbf{x})$ are the unnormalized scores the network generates for each class before applying a softmax function, which maps the logits to the probability distribution over the classes. Logits represent the internal representations of the input data learned by the network. During training, the network's parameters are adjusted to minimize the difference between predicted probabilities and the actual labels of the training data. This optimization process helps the network learn to assign higher logits to the correct classes and lower logits to incorrect classes. For a $k$-class classifier, the output vector of probabilities $y$ is obtained by applying the softmax function to the logits $Z(\textbf{x})$. The final model prediction is the class with the highest probability. Let $y$ be the output vector of probabilities then, $y = \text{softmax}(Z(\textbf{x})) = \frac{\exp(Z(\textbf{x}))}{\sum_{j=1}^{k} \exp(Z_j(\textbf{x}))}$ where $Z(\textbf{x})$ are the logits generated by the CNN for the input image $\textbf{x}$ and $y$ is the resulting vector of class probabilities.


\textbf{Adversarial Attack:} Though highly accurate, DNNs are vulnerable to adversarial attacks that can cause input misclassifications \cite{goodfellow2015explaining,  carlini2017towards, madry2017towards, biggio2018wild, kurakin2018adversarial}. Given a model $F$ and input sample $\textbf{x},y$, the goal of an adversarial evasion attack is to modify the sample $\textbf{x}$ by adding a perturbation such that $F(\textbf{x}) \neq F(\textbf{x}^*)$ and $||\textbf{x}^*-\textbf{x}|| < \epsilon$ where $\epsilon \in R^n$ is the maximum perturbation allowed and $\textbf{x*}$ is the adversarial sample. In targeted attacks, an adversary aims to misclassify the input sample into a \textit{specific} class such that $ F(\textbf{x}^*)=t$ where $t$ is the target label in the label space. In untargeted attacks, an adversary aims to misclassify an input into any other class but the correct one. Untargeted attacks have fewer perturbations than targeted attacks and have better success rates with strong transferability capability \cite{carlini2017towards,liu2016delving}. Adversarial attacks also differ based on the distance measures, usually defined as $L_p$ norm ($L_0$, $L_1$, $L_2$, and $L_\infty$), between the benign and adversarial input. Based on adversary knowledge of the target classifier, attacks can further be classified as black-box (no information), gray-box (partial information), and white-box (complete information) attacks. For example, the Fast Gradient Sign Attack (FGSM) \cite{goodfellow2015explaining} assumes a linear approximation of the network loss function and finds a perturbation by increasing the local linear approximation of the loss. The Basic Iterative Method (BIM) \cite{kurakin2018adversarial} is an iterative version of FGSM where the perturbation is computed multiple times with small steps. Projected Gradient Descent (PGD) \cite{madry2017towards} is also an iterative method similar to BIM. However, unlike BIM, it starts from a random perturbation in the $L_\infty$ ball around the input sample. Auto-PGD attack \cite{croce2020reliable} is a gradient-based adversarial attack that reduces the parameter dependence on step-size of the PGD attack. Carlini and Wagner (CW) \cite{carlini2017towards} attacks comprise a range of attacks that follow an optimization framework similar to L-BFGS \cite{szegedy2013intriguing}. However, they replace the loss function with an optimization problem involving logits $(Z(.))$, instead of using the model prediction.

\textbf{Defense Against Attack:} There are two categories of approaches to adversarial attack mitigation. The first category focuses on improving model robustness against attacks. For example, adversarial training \cite{goodfellow2015explaining} augments natural training data with adversarial samples and performs model training to build a robust classifier that utilizes both original and perturbed samples. However, this method requires information on the generation of adversarial samples, compromises the benign accuracy of the model, and is still susceptible to adversarial attacks like blind-spot attacks \cite{carlini2017towards, zhang2018limitations}. The second category of defense focuses on detecting and rejecting adversarial samples at test time. The detection can be supervised or unsupervised. The supervised detection methods extract features of benign and adversarial samples and train another network for detection \cite{feinman2017detecting, ma2018characterizing, sperl2020dla, wu2021beating}. However, the detection network can also be compromised by adversarial attacks \cite{carlini2017adversarial}. Since supervised approaches require prior knowledge of attacks and the availability of adversarial samples for training, it can be a major limitation. On the other hand, unsupervised detection methods require only benign samples for training. Such methods extract features from benign samples and compute thresholds that measure inconsistency between properties of benign and adversarial samples. For example, Feature Squeezing \cite{xu2017feature} identifies adversarial images by compressing the input space using image filters and comparing its prediction vectors with that of original images using a threshold learned from benign images. MagNet \cite{meng2017magnet} uses denoisers trained on benign images to reconstruct input samples. It assumes that the threshold-based reconstruction error will be smaller for benign images than for adversarial images. While effective on small datasets (MNIST, CIFAR-10), none of these methods perform well on larger images such as ImageNet \cite{carlini2017adversarial}. DNR \cite{sotgiu2020deep} uses features of benign images from different layers of a network to build a detection method but requires training additional N-SVM classifiers with an RBF kernel. NIC \cite{ma2019nic} also extracts the activation distribution of benign images in network layers but builds an additional set of models.

Similar to our work, Roth et al. \cite{roth2019odds} and Hu et al. \cite{hu2019new} use noise as a probe for adversarial detection. Roth et al. \cite{roth2019odds} compute log-odds robustness, the changes in logits between each pair of predicted classes, for a given image and its noisy counterpart. They learn threshold from benign images and use it for adversarial detection. However, Hosseini et al. \cite{hosseini2019odds} generate adversarial images using mimicry attacks that can bypass this statistical approach. The method also requires generating over 2000 noisy samples per image, making it unsuitable when prediction time is of concern. Hu et al. \cite{hu2019new} compare the change in softmax scores between the original and noisy input images and learn the detection threshold from benign images. However, the prediction probability from a softmax distribution poorly corresponds to the model's confidence \cite{hendrycks2016baseline}. Consequently, applying the softmax function to logits results in a loss of discriminating information \cite{aigrain2019detecting} and can be ineffective in detecting adversarial detection. Both approaches aim to preserve the model prediction of benign images and do not account for changes in feature attribution.

\textbf{Explanation Method:} Explanation methods consist of techniques that explain a prediction of a black-box model in a post-hoc manner. Given a trained model and a test input, such methods provide explanations in terms of attribution scores or rules to explain why the model made a certain prediction. Feature attribution, a type of post-hoc explanation method, assigns an attribution score to each input feature of an instance, indicating its importance in the model's prediction \cite{ribeiro2016should}. Given a trained model $F(.)$ and a test instance $\textbf{x}\in R^d$, a feature attribution-based explanation method $\phi$ returns an attribution vector $\phi(\textbf{x}) \in R^d$. The attribution vector is a vector of scores that quantify the importance of the input features in the model prediction of the test instance.  Backpropagation-based attribution methods utilize gradients to propagate the model prediction to the input layer. For example, the Vanilla Gradient method \cite{simonyan2013deep} calculates the gradient of the class score (output) with respect to the input. Integrated Gradient (IG) method \cite{sundararajan2017axiomatic} aggregates gradients along a linear path from a baseline to the test input. The choice of the baseline is specific to the use-case \cite{sturmfels2020visualizing} and should have a near-zero score for model prediction. IG satisfies fundamental axioms of attribution methods: sensitivity (the feature is essential for prediction), implementation (attribution method is efficient and scalable), and completeness (attribution score of input features adds up to output score for an input) \cite{sundararajan2017axiomatic}. It also produces more stable explanations than the Vanilla Gradient method \cite{Bhusal2022SoKME}. 

\textbf{Attack Detection Using Explanation:} Recent research has explored using feature attribution for adversarial detection. For example: (a) Tao et al. \cite{tao2018attacks} identify critical neurons in feature attribution of faces to detect adversarial images. However, their approach is limited to face recognition systems. (b) Zhang et al. \cite{zhang2018detecting} train a supervised classifier using the Vanilla Gradient method-based feature attribution. However, their additional networks for detection can be vulnerable to adversarial attacks. (c) ML-LOO \cite{yang2020ml} extracts feature attribution using the leave-one-out (LOO) method \cite{li2016understanding} and trains several detectors using benign and adversarial images. However, they train several attack-specific detectors by extracting attribution from hidden layers, which can be computationally expensive and may not be scalable to a large number of attacks. (d) X-ensemble \cite{wang2020interpretability} and ExAD \cite{vardhan2021exad} train an ensemble network by extracting feature attribution using different explanation methods for benign and adversarial images. However, this approach also requires prior information on attacks and feature attribution for various adversarial samples and explanation methods, making it difficult to apply in real-world scenarios.

\section{Motivation}\label{sec:motivation}
We provide motivations for our detector design by discussing adversarial perturbation regions around benign samples, and the influence of noise on model prediction and feature attribution. For this discussion, we consider $F$ to be a target image classification model, $\textbf{x}$ is an input image, and $Z(\textbf{x})$ is the logits given by the model. We sample noise $\eta \in \mathcal{N}(0, \sigma^2)$ (where $\sigma^2$ is a hyperparameter) and obtain a noisy version of the input image $\textbf{x}' = \textbf{x}+\eta$. The logits returned by the model $F$ for $\textbf{x}'$ is given by $Z(\textbf{x}')$. We use Integrated Gradient (IG) \cite{sundararajan2017axiomatic} as our attribution method that provides attribution vector $IG^F(\textbf{x})$ and $IG^F(\textbf{x}')$ for the original and noisy sample.

\textbf{Effect of noise on model prediction:} 
DNNs are susceptible to imperceptible adversarial perturbations in an input sample that can change the predicted label of the sample \cite{goodfellow2015explaining,carlini2017towards}. Early explanations for this vulnerability attributed it to ``blind spots'' in the high-dimensional input space, which are low-probability adversarial pockets that make an input vulnerable to adversarial perturbations \cite{szegedy2013intriguing}. Goodfellow et al. \cite{goodfellow2015explaining} explain this vulnerability of neural networks in terms of the linear behavior of networks in high-dimensional spaces. Tanay et al. \cite{tanay2016boundary} demonstrate that adversarial attacks are possible because most input samples in high-dimensional space are bound to exist near the class boundary, making them susceptible to even minor modifications.

Furthermore, it has been shown that by carefully traversing data manifolds, the true label of an input sample can be changed by perturbing it off the data manifold \cite{gardner2015deep}. We can do so by introducing noise (e.g., Gaussian) to an input. Prior works \cite{hu2019new} have modified an input with noise and studied how the model responds to gain insights into the model’s behavior for adversarial detection. Hu et al. \cite{hu2019new} compute softmax scores before and after adding noise to an image and measure the change. They empirically pick a noise parameter that preserves the behavior of benign images on the assumption that natural images are robust to random input corruptions compared with adversarial images. However, such additional noise can amplify the adversarial effects and increase the likelihood of fooling the DNN. This impact depends on the noise parameter and nature of the dataset. For example, for lower-dimension datasets like MNIST, both benign and adversarial images are concentrated in a low-dimension manifold \cite{shafahiadversarial}. While benign images stay robust to noise, adversarial images move off the manifold, producing \textit{significant} changes to the model prediction.

On the contrary, for higher dimensions, benign images tend to lie close to decision boundaries, making them susceptible to small noise that can lead to misclassification \cite{tanay2016boundary}. Adversarial images, on the other hand, often lie in the space outside the data manifold. These are low-probability manifolds created due to a lack of images in the training dataset. Hence, additional noise to benign images can change their position relative to their original position in the manifold, producing \textit{significant} variation in model prediction. However, because the adversarial samples already lie on the low-probability manifold, model sensitivity to additional noise is minimal. This sensitivity of an image to noise can be measured by comparing the change in logits $  \delta_1 = ||Z(\textbf{x}')-Z(\textbf{x})||$.

\begin{figure}[hbt!]
    \centering
    \fbox{\includegraphics[width=.23\textwidth]{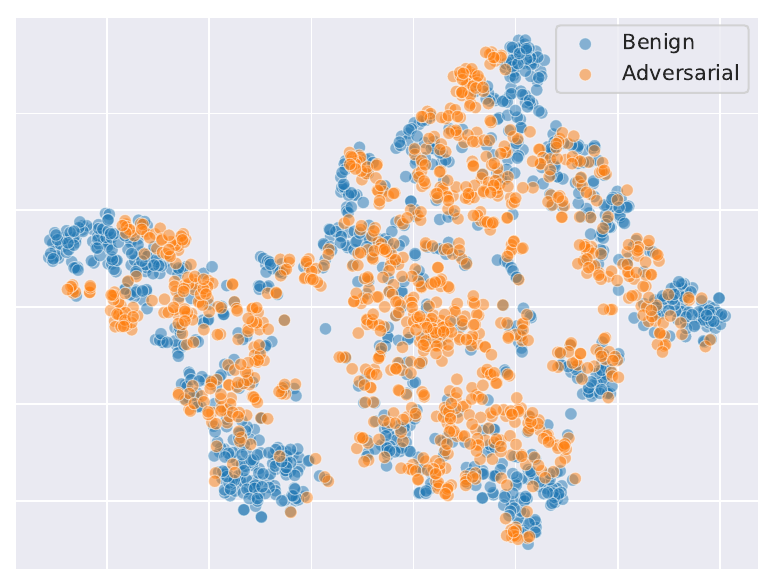}\hfill 
    \includegraphics[width=.23\textwidth]{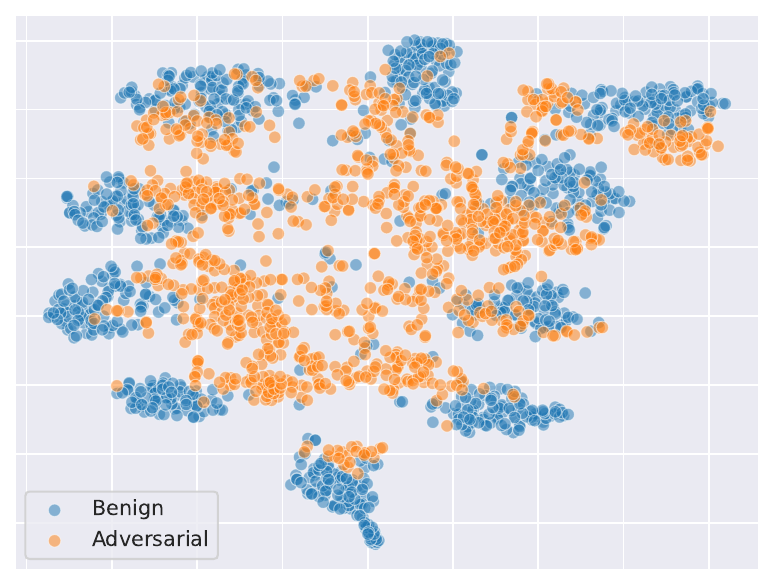}}
    \caption{2D Visualization: Adversarial vs. Benign Classifier Features. Left: Shared manifold for simpler images (MNIST). Right: Adversarial samples form out-of-distribution features in complex images (CIFAR-10).}
    \label{fig:inputSpaceViz}
\end{figure}

\textbf{Illustrative Example:} We project the feature extracted from the penultimate layer of the trained CNN classifiers into a 2D space using the t-SNE algorithm \cite{van9visualizing} (see Figure \ref{fig:inputSpaceViz}). We use 10000 representative benign images and their adversarial counterpart with an untargeted $L_\infty$ FGSM attack $\epsilon=8/255$ from the training set of each dataset. For MNIST, the benign images and their adversarial counterparts lie in a low-dimensional manifold, suggesting that adversarial images are crafted by traversing the data manifold observed in the training distribution. However, for CIFAR-10, we observe that adversarial images primarily lie outside the distribution observed in the training data. This suggests that adversarial samples for complex images are created by introducing out-of-distribution samples where network behavior is not predictable based on the training data. Given that benign images lie near the decision boundary, adding sufficient noise can push them to the out-of-distribution region, thus resulting in a significant change in model predictions than their adversarial counterparts.

\textbf{Effect of noise on feature attribution:} Feature attribution-based explanation methods assign a score to each input feature, indicating their importance in the model's prediction for an instance. The distribution of such feature attribution scores varies for adversarial and benign samples and can be measured using statistical measures of dispersion \cite{yang2020ml}. Figure \ref{fig:advSample} shows heat maps for benign and adversarial counterparts for the ImageNet samples using the IG method, highlighting the contrast in attribution distribution. We observe that positive attribution scores (red pixels) are distributed across more input features in adversarial images than in benign images. This observation underscores the sensitivity of gradient-based feature attribution methods to perturbations in the input. 

\textbf{Relationship between IG sensitivity and model sensitivity:} The feature attribution score computed by IG for feature $i$ of input sample $\textbf{x} \in R^d$ with baseline $\textbf{u}$, model $F$ is given by:
\begin{equation}\label{eqn:igequationsimple}
    IG_i^F (\textbf{x,u}) = (x_i - u_i). \int_{\alpha=0}^{1} {\partial_i F(\textbf{u}+\alpha (\textbf{x} - \textbf{u}))} \partial \alpha 
\end{equation}

For an input sample $\textbf{x}$, IG returns a vector $IG^F(\textbf{x}, \textbf{u}) \in R^d$ with scores that quantify the contribution of $x_i$ to the model prediction $F(\textbf{x})$. For a single layer network $F(\textbf{x}) = H(<\textbf{w},\textbf{x}>)$ where $H$ is a differentiable scalar-valued function and $<\textbf{w},\textbf{x}>$ is the dot product between the weight vector $\textbf{w}$ and input $\textbf{x}\in R^d$, IG attribution has a closed form expression \cite{chalasani2020concise}.

For given \textbf{x}, \textbf{u} and $\alpha$, let us consider $\textbf{v}= \textbf{u}+\alpha (\textbf{x}-\textbf{u})$. If the single-layer network is represented as $F(\textbf{x}) = H(<\textbf{w},\textbf{x}>)$ where $H$ is a differentiable scalar-valued function, $\partial_iF(\textbf{v})$ can be computed as:

\begin{align}\label{eqn:partialfv}
    \partial_i F(\textbf{v}) &= \frac{\partial F(\textbf{v})}{v_i} = \frac{\partial H(<\textbf{w}, \textbf{v}>)}{\partial v_i} = H'(z) \frac{\partial<\textbf{w}, \textbf{v}>}{\partial v_i} \notag\\ 
    &= w_i H'(z)  
\end{align}

Here, $H'(z)$ is the gradient of the activation $H(z)$ where $z = <\textbf{w}, \textbf{v}>$. To compute $\frac{\partial F(\textbf{v})}{\partial \alpha}$:

\begin{equation}\label{eqn:computeto}
    \frac{\partial F(\textbf{v})}{\partial \alpha} = \sum_{i=1}^d (\frac{\partial F(\textbf{v})}{\partial v_i} \frac{\partial v_i}{\partial \alpha})
\end{equation}

We can substitute value of $ \frac{\partial v_i}{\partial \alpha} = (x_i-u_i)$ and $\partial_i F(\textbf{v})$ from Eq. \ref{eqn:partialfv} to Eq. \ref{eqn:computeto}.

\begin{align}
    \frac{\partial F(\textbf{v})}{\partial \alpha} &= \sum_{i=1}^d [w_i H'(z) (x_i - u_i)] \notag \\  
    &= <\textbf{x}-\textbf{u}, \textbf{w}>H'(z)
\end{align}

This gives: 
\begin{equation}
    dF(\textbf{v}) = <\textbf{x}-\textbf{u}, \textbf{w}> H'(z)\partial\alpha 
\end{equation}

Since $<\textbf{x}-\textbf{u}, \textbf{w}>$ is scalar, 
\begin{equation}\label{eqn:torewriteIG}
    H'(z)\partial\alpha = \frac{dF(\textbf{v})}{<\textbf{x}-\textbf{u}, \textbf{w}>} 
\end{equation}

Eq. \ref{eqn:torewriteIG} can be used to rewrite the integral in the definition of $IG_i^F(\textbf{x})$ in Eq. \ref{eqn:igequationsimple},

\begin{align}
    \int_{\alpha=0}^{1} \partial_i F(\textbf{v})\partial\alpha &= \int_{\alpha=0}^{1} w_i H'(z) \partial z \notag ~~~ \textnormal{[From Eqn. \ref{eqn:partialfv}]}\\ 
    &= \int_{\alpha=0}^{1} w_i \frac{ dF(\textbf{v})}{<\textbf{x}-\textbf{u}, \textbf{w}>} \notag\\ 
     &= \frac{w_i}{<\textbf{x}-\textbf{u}, \textbf{w}>} \int_{\alpha=0}^{1}{d F(\textbf{v})} \notag \\
     &= \frac{w_i}{<\textbf{x}-\textbf{u}, \textbf{w}>} [F(\textbf{x})-F(\textbf{u})]
\end{align}

Hence, we obtain the closed form for IG from its definition in Eqn. \ref{eqn:igequationsimple} as

\begin{align}\label{eqn:igequation}
    IG_i^F(\textbf{x},\textbf{u}) &= [F(\textbf{x}) - F(\textbf{u})] \frac{({x_i} - {u_i}){w_i}}{<\textbf{x} - \textbf{u}, \textbf{w}>} \notag \\
    IG^F(\textbf{x},\textbf{u})  &= [F(\textbf{x}) - F(\textbf{u})] \frac{(\textbf{x} - \textbf{u}) \odot
 \textbf{w}}{<\textbf{x} - \textbf{u}, \textbf{w}>}
\end{align}

Here, $\odot$ is the entry-wise produce of two vectors. 

Eq. \ref{eqn:igequation} shows that the feature attribution in IG is proportional to the fractional contribution of a feature to the change in logit $<\textbf{x} - \textbf{u}, \textbf{w}>$. When an adversary perturbs an input sample for changing the predicted label, the value of logits changes accordingly. In untargeted attacks, the adversarial perturbation aims to maximize the softmax value of a class different than the original class. Hence, the perturbation can increase or decrease logit values of other classes \cite{aigrain2019detecting}. This change in logits also brings a change in feature attribution. When an additional noise is introduced to an input sample, the change in feature attribution follows the change in model prediction. This sensitivity of IG to noise can be measured using Eq. \ref{eqn:motivationatr}.

\begin{align}\label{eqn:motivationatr}
      \delta_2 &= ||IG^F(\textbf{x}',\textbf{u}) - IG^F(\textbf{x},\textbf{u})||_1 \approx ||IG^F(\textbf{x}',\textbf{x}) ||_1 \notag \\
   &\approx\Big{|}\Big{|}[F(\textbf{x}') - F(\textbf{x})] \frac{(\textbf{x}' - \textbf{x}) \odot
 \textbf{w}}{<\textbf{x}' - \textbf{x}, \textbf{w}>}\Big{|}\Big{|}_1  \notag \\ 
 &\approx \Big{|}\Big{|}[F(\textbf{x}') - F(\textbf{x})] \frac{\Delta \odot
 \textbf{w}}{<\Delta, \textbf{w}>}\Big{|}\Big{|}_1 
\end{align}

\noindent {Assuming $\textbf{w}$ to be constant for a given model, we can conclude from Eqn. \ref{eqn:motivationatr} that $\delta_2 \propto ||F(\textbf{x}')-F(\textbf{x})||$. This implies that the sensitivity of IG is tied to the overall sensitivity of the model. Based on these observations, we posit that the sensitivity of IG could serve as a valuable tool in identifying adversarial samples by providing an additional layer of insight into the behavior of deep learning models.

\begin{figure}[]
    \centering
    \includegraphics[width=.45\textwidth]{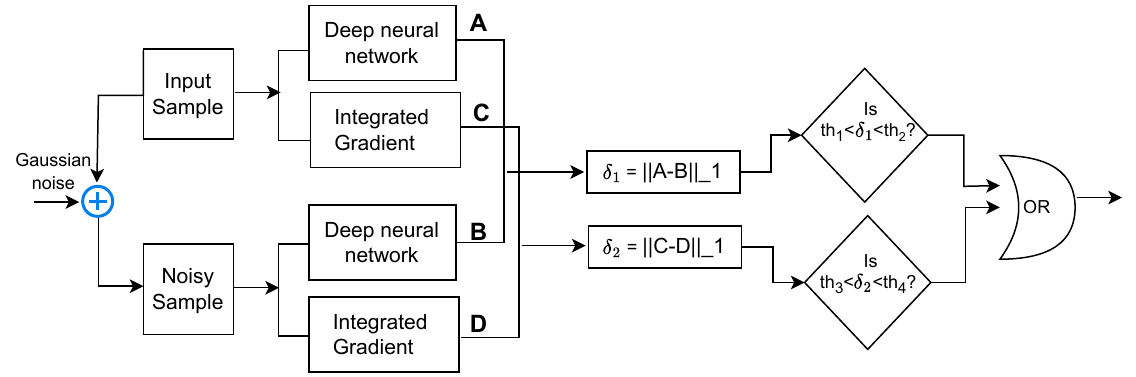}
    \caption{PASA overview: A \& B are neural network outputs (logits), C \& D are IG feature attributions.}
    \label{fig:diagram}
\end{figure}

\section{Methodology}\label{sec:methodology}
\subsection{Threat model}
We consider a classification task with a distribution $D$ over input samples $\textbf{x} \in R^n$ with labels $y \in [K]$. A classifier is a function $F:R^n \rightarrow [K]$ learned by a neural network architecture in a supervised manner that classifies a given input sample into one of $k$ classes. An adversary can manipulate the sample at test time by adding $L_\infty$ perturbation so that the new sample $\textbf{x}^*$ is an adversarial sample and wrongly classified by the classifier. A detector $f_{det}$ is a function that computes a score for the given input sample based on our proposed approach and decides whether the sample is benign or adversarial by comparing it against a learned threshold. The optimal threshold for each dataset is learned during the training phase of the detector  (See Section \ref{sec:design}). At test time, we assume no previous knowledge of the underlying attack mechanism. Below, we describe the set of assumptions about an adversary for our proposed method and its evaluation.

\subsubsection{Adversary goal}
Adversarial samples are inputs specifically designed to produce targeted or untargeted misclassification from a targeted machine learning model. We assume that the adversary is not concerned with a specific target label and only aims to produce misclassification. Untargeted attacks require fewer perturbations than targeted attacks and are more difficult to detect \cite{carlini2017towards}.

\subsubsection{Adversary capabilities} 

Defenses to adversarial samples typically restrict the adversary's capability to make ``small" changes to the given input. In the case of image classification, this change is measured in $L_p$ norm between two inputs for $p \in [0,1,2, \infty]$. We assume that the adversary performs $L_\infty$ attack with the constraint of $\epsilon$, which means that the attack cannot modify the pixel of the input image by more than $\epsilon$ \cite{goodfellow2015explaining}. However, we evaluate our results on different $\epsilon$ specifically, $\epsilon \in [8/255,16/255,32/255,64/255]$. 

\subsubsection{Adversary knowledge}
We evaluate our detector under white-box assumption where an adversary has complete knowledge of the target model and its parameters and dataset used for training. We perform two categories of white-box attacks: (a) an adversary has access to the model so that it can create an attack to evade the classification; (b) in addition to the target model, an adversary has knowledge of the underlying detector and can modify their attack to evade target model, and the detection mechanism.

\subsection{Proposed design}\label{sec:design}
 Based on our insights on the sensitivity of model prediction and feature attribution (discussed in Section \ref{sec:motivation}), we propose using noise as a probing mechanism for adversarial detection. The underlying principle is that the characteristics of model prediction and feature attribution on the noise-corrupted sample differ depending on whether the sample is natural or adversarial. We add noise to a sample and measure the change in model prediction and feature attribution. Our detector (see Figure \ref{fig:diagram}) classifies an input sample as adversarial if the change in either the model prediction or feature exceeds a learned threshold established from benign samples. 

\begin{figure}[]
    \centering
    \includegraphics[width=.25\textwidth]{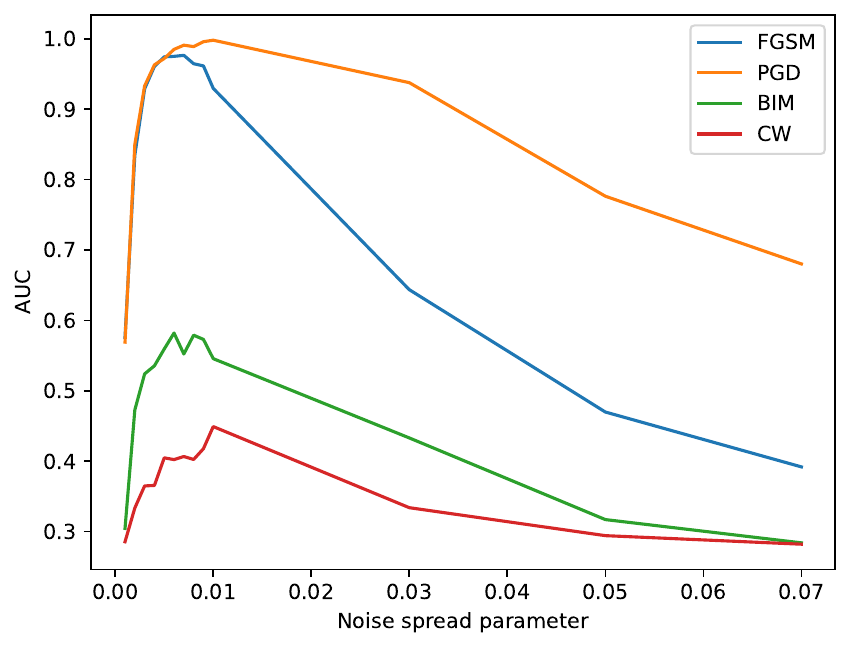}
    \caption{Performance of PASA for MNIST against various adversarial attacks at varying noise spread parameters.}
    \label{fig:parameter}
\end{figure}

\textbf{Training:} Given a black box model $F(\textbf{.})$, and an input sample $\textbf{x}$, the model outputs logits, $Z(\textbf{x})$. Feature attribution method, Integrated Gradient (IG), gives attribution vector $IG^F(\textbf{x})$. To derive a noisy version of the input sample, we add Gaussian noise $\eta \in \mathcal{N}(0, \sigma^2)$, where $\sigma^2$ is a hyperparameter and equals $(max(\textbf{x})-min(\textbf{x}))*spread$. The noisy sample ($\textbf{x}'$) is obtained as $\textbf{x}+\eta$. $spread$ controls the standard deviation of the noise and is our only hyper-parameter required for detector design. We vary the parameter \textit{spread} under different values for each dataset and empirically select the value that gives us the best adversarial detection performance.  For example, Figure \ref{fig:parameter} shows the performance of our detector on various noise spread parameters for the MNIST dataset with different adversarial attacks at $\epsilon=0.15$. We can observe that the detector has the maximum AUC at the noise spread parameter 0.005. We followed the same procedure on updated-CIC-IDS2017, CIFAR-10, CIFAR-100, and ImageNet and obtained the noise-spread parameter as 0.0005, 0.15, 0.15, and 0.35 respectively.

Next, we compute the logit and feature attribution of the noisy sample (\textbf{x}') and measure the change using the $L_1$ norm of the difference. We term these changes as \textit{prediction sensitivity (PS)}, and \textit{attribution sensitivity (AS)} as expressed in Eq. \ref{eqn:robustnessmodelpred} and Eq \ref{eqn:deltaattributionrobustness} respectively. 

\begin{equation}\label{eqn:robustnessmodelpred}
    \delta_1 = ||Z(\textbf{x}') - Z(\textbf{x})||_1 
\end{equation}

\begin{equation}\label{eqn:deltaattributionrobustness}
    \delta_2 = ||IG^F(\textbf{x}', \textbf{u})-IG^F(\textbf{x}, \textbf{u})||_1
\end{equation} 

We demonstrate the different characteristics of model prediction and feature attribution on noise-corrupted images for MNIST and CIFAR-10 in Figure \ref{fig:robustnessPredGaussian} and Figure \ref{fig:featureAttrGaussian}. As explained in training, we first collect benign and adversarial images of both datasets, add Gaussian noise (spread parameter of 0.005 for MNIST and 0.15 for CIFAR-10), and measure prediction sensitivity and attribution sensitivity. Figure \ref{fig:robustnessPredGaussian} shows the histogram plots for a set of benign and adversarial image prediction sensitivity. For MNIST, benign samples demonstrate smaller norms compared to their adversarial counterparts, indicating that they can be distinguished from adversarial samples with a threshold range $[0-3]$. This behavior is true for datasets like MNIST, where images are concentrated in low-dimensional manifolds. In contrast, for a three-channel image dataset like CIFAR-10, we observe a \textit{divergent} behavior. The difference in model prediction for noisy and original images in benign samples is greater than that of adversarial samples and their noisy counterparts. This behavior is due to the distinct positions benign and adversarial images occupy within the input space manifold, as discussed in Section \ref{sec:motivation}. We can also observe that for CIFAR-10, adversarial samples generated with a larger perturbation parameter ($\epsilon$) exhibit minimal changes in model prediction. This is because the adversarial images are located far from the decision boundary, and the added noise has minimal impact. ImageNet and CIFAR-100 demonstrate similar behavior.

\begin{figure}[]
    \centering
    \includegraphics[width=.23\textwidth]{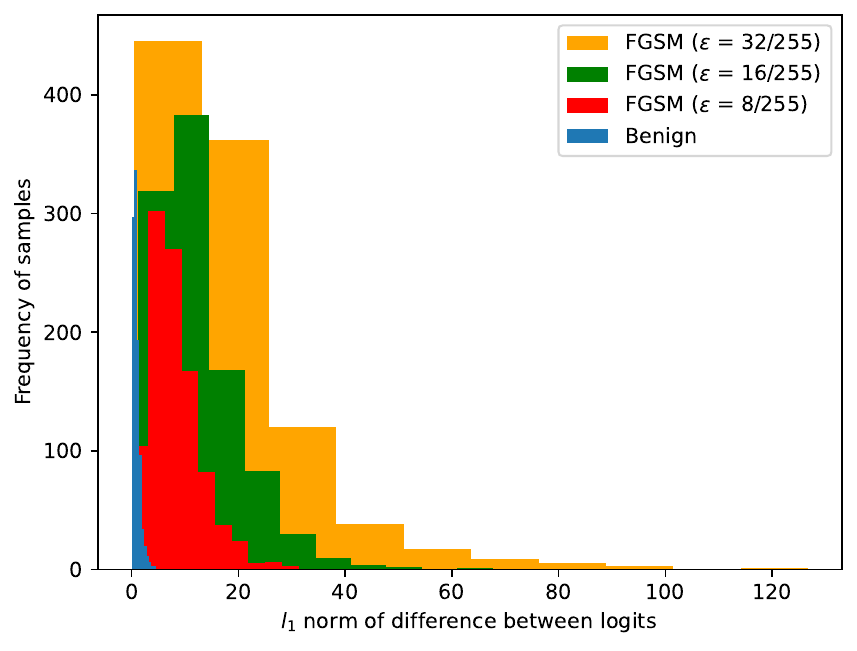}\hfill 
    \includegraphics[width=.23\textwidth]{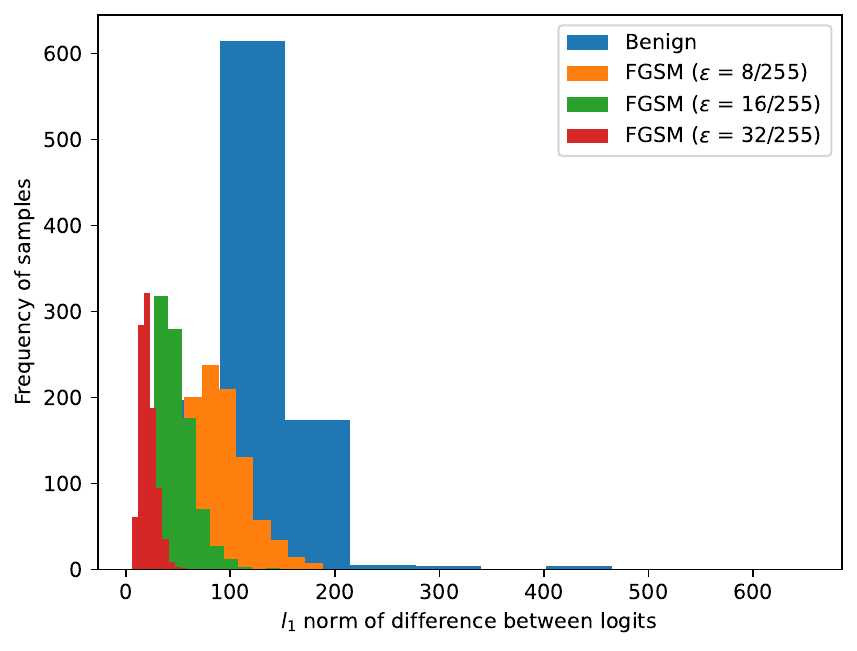}
    \caption{The distribution of the difference between logits of benign and adversarial images with their noisy counterparts (left: MNIST, right: CIFAR-10). Adversarial samples are obtained at various perturbation strengths $\epsilon$. 
    }
    \label{fig:robustnessPredGaussian}
\end{figure}

\begin{figure}[]
    \centering
    \includegraphics[width=.23\textwidth]{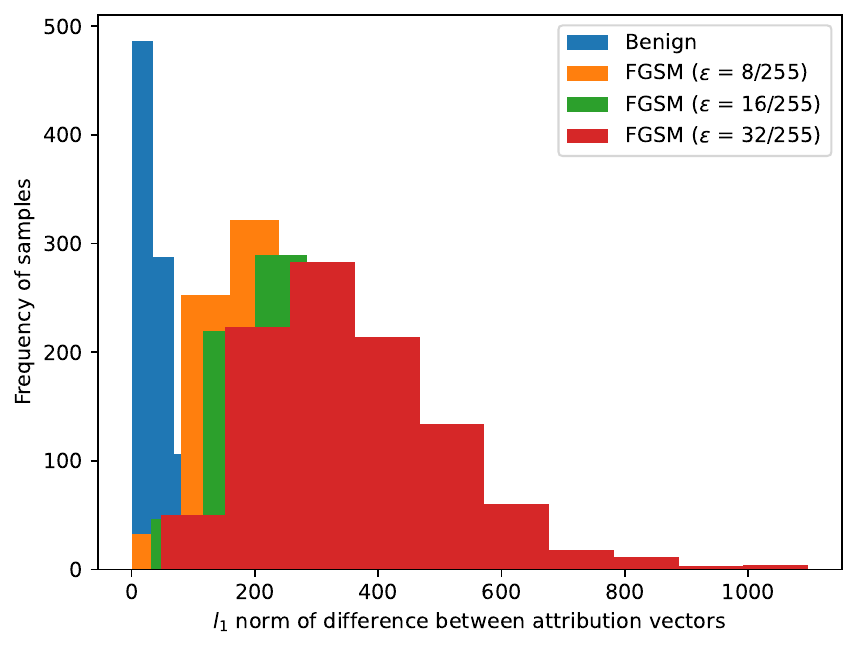}\hfill 
    \includegraphics[width=.23\textwidth]{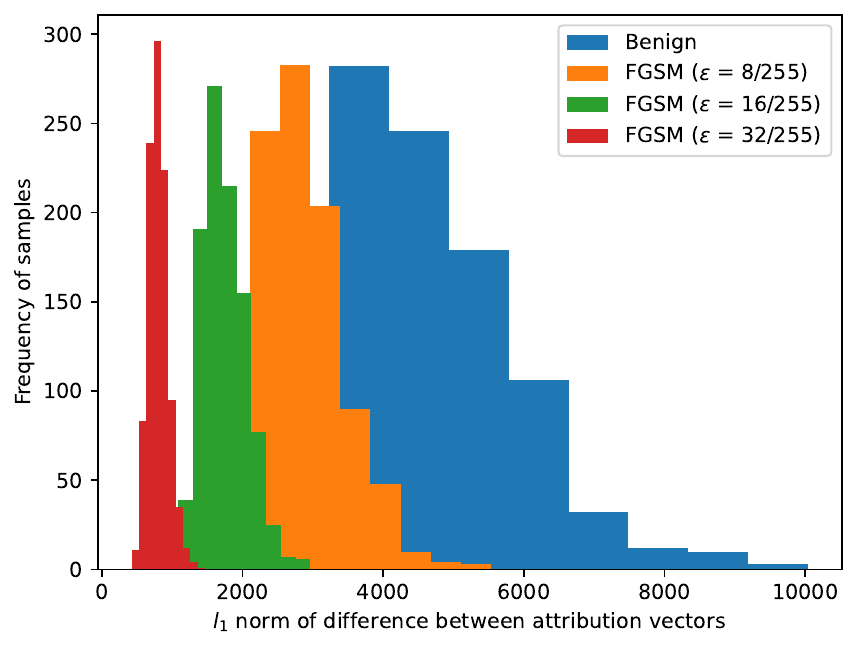}
    \caption{The distribution of the difference between the attribution vector of benign and adversarial images with their noisy counterparts (left: MNIST, right: CIFAR-10). Adversarial samples are obtained at various perturbation strengths $\epsilon$. 
    }
    \label{fig:featureAttrGaussian}
\end{figure}

Figure \ref{fig:featureAttrGaussian} shows the histogram plots for a set of benign and adversarial images of the MNIST and CIFAR-10 datasets for attribution sensitivity. We observed contrasting model prediction sensitivity between MNIST and CIFAR-10 in Figure \ref{fig:robustnessPredGaussian}. Since feature attribution of an image relies on the model prediction as demonstrated by Eq. \ref{eqn:igequation}, the feature attribution sensitivity distribution follows the model prediction behavior. While for MNIST, the benign and its noisy counterparts have a smaller $L_1$ norm, the opposite is true for CIFAR-10. ImageNet and CIFAR-100 demonstrate similar behavior.

Training PASA thus involves learning the threshold for the prediction sensitivity and attribution sensitivity for benign samples. For each dataset, we collect 5000 benign samples from the training set, probe them with noise, measure the prediction sensitivity and attribution sensitivity metrics, and learn the threshold that yields various false positive rates (FPRs) on a validation set of benign samples. We provide the methodology of our approach below: 
\newline \noindent 
\textbf{Methodology:} 
\newline \noindent 
\textit{Step 1.} Set noise spread parameter $\sigma$ as 0.001 for MNIST, 0.0001 for CIC-IDS2017, 0.1 for CIFAR, and ImageNet.
\newline \noindent
\textit{Step 2.} For a set of benign samples, produce its noisy version by adding Gaussian noise. Compute two metrics, PS and AS (See Eqns \ref{eqn:robustnessmodelpred} \& \ref{eqn:deltaattributionrobustness}).
\newline \noindent 
\textit{Step 3.} Find thresholds of PS and AS that produce 1\%, 5\%, and 10\% False Positive Rate (FPR) on a hold-out set of benign samples.
\newline \noindent    
\textit{Step 4.} Evaluate the detection results on a validation set (consisting of benign and adversarial samples) using the threshold and noise parameter learned from Step 3.
\newline \noindent 
\textit{Step 5.} Increment the noise to $\sigma' = \sigma + \delta$, where $\delta$ is dataset-dependent. The following delta levels worked best in our experiment: 0.01 for MNIST, 0.0001 for CIC-IDS2017, and 0.1 for CIFAR and ImageNet.
\newline \noindent 
\textit{Step 6.} Repeat Steps 2-5.
\newline \noindent
\textit{Step 7.} Pick the best-performing noise spread parameter and threshold.

\textbf{Testing:} At test time, we evaluate changes in model prediction and feature attribution of an input sample. We add Gaussian noise with zero mean and standard deviation of $(max(\textbf{x})-min(\textbf{x}))*spread$, where we select spread empirically during training. We compute prediction sensitivity and attribution sensitivity using expressions of Eq. \ref{eqn:robustnessmodelpred} and Eq. \ref{eqn:deltaattributionrobustness}. We reject a sample as adversarial if either of the computed metrics does not satisfy the threshold learned during training.

\begin{table*}[]
\centering
\caption{Adversarial Detection Performance for MNIST and CIFAR-10 models: Our Method (PASA) vs. Unsupervised Methods (FS, MagNet, U-LOO, TWS) using AUC scores.}
\label{tab:mnistcifarresult}
\resizebox{\textwidth}{!}{%
\begin{tabular}{@{}lclllll|lllll|lllll@{}}
\toprule
\textbf{Attack} &
  \multicolumn{1}{l}{\textbf{}} &
  \multicolumn{5}{c|}{\textbf{MNIST}} &
  \multicolumn{5}{c|}{\textbf{CIFAR-10 (VGG)}} &
  \multicolumn{5}{c}{\textbf{CIFAR-10 (ResNet)}} \\ \midrule
\textbf{} &
  \textbf{Strength} &
  \multicolumn{1}{c}{\textbf{FS}} &
  \multicolumn{1}{c}{\textbf{MagNet}} &
  \multicolumn{1}{c}{\textbf{U-LOO}} &
  \multicolumn{1}{c}{\textbf{TWS}} &
  \multicolumn{1}{c|}{\textbf{PASA}} &
  \multicolumn{1}{c}{\textbf{FS}} &
  \multicolumn{1}{c}{\textbf{MagNet}} &
  \multicolumn{1}{c}{\textbf{U-LOO}} &
  \multicolumn{1}{c}{\textbf{TWS}} &
  \multicolumn{1}{c|}{\textbf{PASA}} &
  \multicolumn{1}{c}{\textbf{FS}} &
  \multicolumn{1}{c}{\textbf{MagNet}} &
  \multicolumn{1}{c}{\textbf{U-LOO}} &
  \multicolumn{1}{c}{\textbf{TWS}} &
  \multicolumn{1}{c}{\textbf{PASA}} \\ \midrule
\textbf{FGSM} &
  \textbf{8/255} &
  0.89$\pm$0.01 &
  0.94$\pm$0.01 &
  0.93$\pm$0.01 &
  0.94$\pm$0.01 &
  \textbf{0.97$\pm$0.01} &
  0.57$\pm$0.01 &
  0.62$\pm$0.01 &
  0.52$\pm$0.01 &
  0.51$\pm$0.01 &
 \textbf{0.63$\pm$0.02 }&
  0.76$\pm$0.01 &
  0.63$\pm$0.01 &
  0.55$\pm$0.01 &
  0.72$\pm$0.01 &
  \textbf{0.87$\pm$0.01} \\
\textbf{} &
  \textbf{16/255} &
  0.87$\pm$0.01 &
  0.95$\pm$0.01 &
  0.92$\pm$0.01 &
  0.93$\pm$0.02 &
 \textbf{0.98$\pm$0.01}&
  0.68$\pm$0.02 &
  \textbf{0.82$\pm$0.01} &
  0.52$\pm$0.02 &
  0.66$\pm$0.02 &
  0.77$\pm$0.02 &
  0.81$\pm$0.01 &
  0.83$\pm$0.01 &
  0.53$\pm$0.03 &
  0.78$\pm$0.02 &
  \textbf{0.97$\pm$0.01} \\
\textbf{} &
  \textbf{32/255} &
  0.86$\pm$0.01 &
  0.95$\pm$0.01 &
  0.91$\pm$0.02 &
  0.89$\pm$0.04 &
  \textbf{0.98$\pm$0.01} &
  0.66$\pm$0.01 &
  \textbf{0.96$\pm$0.04} &
  0.52$\pm$0.01 &
  0.64$\pm$0.01 &
  0.88$\pm$0.01 &
  0.76$\pm$0.01 &
  0.94$\pm$0.01 &
  0.60$\pm$0.02 &
  0.76$\pm$0.01 &
  \textbf{0.98$\pm$0.01} \\
\textbf{} &
  \textbf{64/255} &
  0.86$\pm$0.01 &
  0.95$\pm$0.01 &
  0.88$\pm$0.02 &
  0.82$\pm$0.10 &
  \textbf{0.98$\pm$0.02} &
  0.63$\pm$0.02 &
  0.95$\pm$0.05 &
  0.49$\pm$0.03 &
  0.61$\pm$0.01 &
  \textbf{0.95$\pm$0.01} &
  0.84$\pm$0.02 &
  0.95$\pm$0.03 &
  0.53$\pm$0.03 &
  0.83$\pm$0.01 &
  \textbf{0.98$\pm$0.02} \\ \midrule
\textbf{PGD} &
  \textbf{8/255} &
  0.90$\pm$0.01 &
  0.95$\pm$0.03 &
 \textbf{0.99$\pm$0.01} &
  0.92$\pm$0.11 &
  0.98$\pm$0.01 &
  0.52$\pm$0.01 &
  0.59$\pm$0.03 &
  0.49$\pm$0.05 &
  0.58$\pm$0.02 &
  \textbf{0.74$\pm$0.02} &
  0.25$\pm$0.01 &
  0.57$\pm$0.04 &
  0.62$\pm$0.01 &
  0.14$\pm$0.01 &
  \textbf{0.83$\pm$0.03} \\
 &
  \textbf{16/255} &
  0.88$\pm$0.01 &
  0.95$\pm$0.04 &
  \textbf{0.99$\pm$0.01} &
  0.76$\pm$0.10 &
  0.98$\pm$0.01 &
  0.59$\pm$0.01 &
  0.75$\pm$0.02 &
  0.51$\pm$0.03 &
  0.56$\pm$0.01 &
  \textbf{0.82$\pm$0.01} &
  0.16$\pm$0.01 &
  0.73$\pm$0.03 &
  0.65$\pm$0.01 &
  0.14$\pm$0.02 &
  \textbf{0.93$\pm$0.02} \\
 &
  \textbf{32/255} &
  0.77$\pm$0.02 &
  0.94$\pm$0.04 &
  0.99$\pm$0.02 &
  0.32$\pm$0.03 &
  \textbf{0.99$\pm$0.01} &
  0.62$\pm$0.02 &
  \textbf{0.93$\pm$0.03} &
  0.51$\pm$0.01 &
  0.53$\pm$0.01 &
  0.90$\pm$0.02 &
  0.13$\pm$0.01 &
  0.91$\pm$0.05 &
  0.68$\pm$0.02 &
  0.14$\pm$0.03 &
  \textbf{0.98$\pm$0.01} \\
 &
  \textbf{64/255} &
  0.48$\pm$0.01 &
  0.95$\pm$0.03 &
  0.99$\pm$0.03 &
  0.11$\pm$0.01 &
  \textbf{0.99$\pm$0.02} &
  0.60$\pm$0.01 &
  0.95$\pm$0.05 &
  0.52$\pm$0.03 &
  0.51$\pm$0.04 &
  \textbf{0.95$\pm$0.01} &
  0.16$\pm$0.01 &
  0.95$\pm$0.06 &
  0.72$\pm$0.02 &
  0.14$\pm$0.02 &
  \textbf{0.98$\pm$0.02} \\ \midrule
\textbf{BIM} &
  \textbf{8/255} &
  \textbf{0.89$\pm$0.03} &
  0.83$\pm$0.01 &
  0.40$\pm$0.04 &
  0.83$\pm$0.02 &
  0.62$\pm$0.04 &
  0.37$\pm$0.02 &
  0.54$\pm$0.02 &
  0.50$\pm$0.01 &
  0.55$\pm$0.05 &
  \textbf{0.66$\pm$0.02} &
  0.29$\pm$0.02 &
  0.53$\pm$0.03 &
  0.60$\pm$0.02 &
  0.16$\pm$0.01 &
  \textbf{0.75$\pm$0.01} \\ 
 &
  \textbf{16/255} &
  0.88$\pm$0.01 &
  \textbf{0.93$\pm$0.01} &
  0.67$\pm$0.04 &
  0.92$\pm$0.01 &
  0.58$\pm$0.03 &
  0.16$\pm$0.01 &
  0.60$\pm$0.04 &
  0.51$\pm$0.01 &
  0.55$\pm$0.06 &
  \textbf{0.71$\pm$0.01} &
  0.16$\pm$0.02 &
  0.58$\pm$0.03 &
  0.59$\pm$0.01 &
  0.15$\pm$0.01 &
  \textbf{0.84$\pm$0.01} \\
 &
  \textbf{32/255} &
  0.88$\pm$0.01 &
  \textbf{0.95$\pm$0.01} &
  0.92$\pm$0.04 &
  0.85$\pm$0.01 &
  0.56$\pm$0.02 &
  0.15$\pm$0.02 &
  0.73$\pm$0.06 &
  0.53$\pm$0.02 &
  0.55$\pm$0.04 &
  \textbf{0.73$\pm$0.02} &
  0.13$\pm$0.01 &
  0.72$\pm$0.04 &
  0.58$\pm$0.01 &
  0.14$\pm$0.02 &
  \textbf{0.93$\pm$0.01} \\
 &
  \textbf{64/255} &
  0.88$\pm$0.01 &
  0.95$\pm$0.02 &
  \textbf{0.99$\pm$0.02} &
  0.69$\pm$0.02 &
  0.55$\pm$0.01 &
  0.13$\pm$0.01 &
  \textbf{0.91$\pm$0.01} &
  0.52$\pm$0.02 &
  0.54$\pm$0.01 &
  0.74$\pm$0.01 &
  0.12$\pm$0.01 &
  0.90$\pm$0.04 &
  0.57$\pm$0.01 &
  0.14$\pm$0.01 &
  \textbf{0.97$\pm$0.02} \\ \midrule
\textbf{Auto-PGD} &
  \textbf{0.15} &
  0.81$\pm$0.02 &
  0.95$\pm$0.01 &
  0.98$\pm$0.03 &
  0.56$\pm$0.03 &
  \textbf{0.98$\pm$0.01} &
  0.14$\pm$0.02 &
  0.96$\pm$0.01 &
  0.52$\pm$0.04 &
  0.12$\pm$0.02 &
  \textbf{0.97$\pm$0.01} &
  0.13$\pm$0.01 &
  0.95$\pm$0.01 &
  0.75$\pm$0.01 &
  0.13$\pm$0.01 &
  \textbf{0.98$\pm$0.02} \\\midrule
\textbf{CW} &
  \textbf{0.15} &
  0.88$\pm$0.03 &
  0.94$\pm$0.02 &
  0.91$\pm$0.02 &
  \textbf{0.95$\pm$0.02} &
  0.58$\pm$0.02 &
  0.66$\pm$0.01 &
  0.71$\pm$0.03 &
  0.54$\pm$0.04 &
  0.68$\pm$0.01 &
  \textbf{0.82$\pm$0.01} &
  0.84$\pm$0.01 &
  0.93$\pm$0.01 &
  0.55$\pm$0.03 &
  0.82$\pm$0.01 &
  \textbf{0.98$\pm$0.01} \\ \midrule
\textbf{Average} &
  \multicolumn{1}{l}{} &
  0.84$\pm$0.10 &
  \textbf{0.94$\pm$0.03} &
  0.90$\pm$0.16 &
  0.75$\pm$0.25 &
  0.83$\pm$0.20 &
  0.46$\pm$0.21 &
  0.79$\pm$0.15 &
  0.51$\pm$0.01 &
  0.54$\pm$0.13 &
  \textbf{0.80$\pm$0.11} &
  0.40$\pm$0.31 &
  0.79$\pm$0.16 &
  0.61$\pm$0.08 &
  0.37$\pm$0.31 &
  \textbf{0.93$\pm$0.07} \\ \bottomrule
\end{tabular}%
}
\end{table*}

\section{Experiment and Evaluation}\label{sec:experiment}
\subsection{Experiment Setup}

We implemented PASA using Python and PyTorch and conducted experiments on a server with a 4 Intel(R) Core(TM) i5-7600K CPU @ 3.80 GHz and a 12 GB NVIDIA TITAN Xp GPU card. We used Captum \cite{kokhlikyan2020captum} to generate explanations.

\subsubsection{{Datasets}}

We evaluate the performance of PASA on the following datasets: MNIST \cite{lecun1998gradient},  CIFAR-10 \cite{krizhevsky2009learning}, CIFAR-100 \cite{krizhevsky2009learning}, ImageNet \cite{deng2009imagenet} and updated CIC-IDS2017 \cite{engelen2021troubleshooting}. The datasets are publicly available, and none of them contain personally identifiable information. Details on the dataset can be found in Appendix \ref{appendix:dataset}. 

\subsubsection{{Target networks}}
To demonstrate the generalization of our approach, we evaluate our results by performing adversarial attacks and detection on a variety of networks: MLP \cite{rumelhart1985llearning}, LeNet \cite{lecun1998gradient},  VGG-16 \cite{simonyan2014very}, ResNet \cite{he2016deep}, and MobileNet \cite{sandler2018mobilenetv2}. Details on model architecture can be found in Appendix \ref{appendix:models}.

\subsubsection{Attacks}\label{sec:attackdetail} We evaluate the performance of PASA against inputs perturbed using the following untargeted $L_\infty$ attacks: FGSM \cite{goodfellow2015explaining}, BIM \cite{kurakin2018adversarial} (10 iterations) and PGD \cite{madry2017towards} (step-size $\alpha=\epsilon/10$, 40 iterations) with increasing value of attack parameter $\epsilon \in [8/255, 16/255, 32/255, 64/255] $, Auto-PGD \cite{croce2020reliable} ($\epsilon=0.15$) and zero confidence CW attack \cite{carlini2017towards} ($\epsilon=0.15$, learning rate= 0.01). Adversarial attacks are performed on the test set which is not used for learning the threshold of PASA. Further details are provided in Appendix \ref{appendix:attack}.

\subsection{Evaluation}
\subsubsection{{Baselines}}
We present the experimental evaluation of PASA by comparing its results against four types of unsupervised detectors that use different statistical approaches for adversarial detection. We discuss their implementation in Appendix \ref{appendix:implementation}. 

\underline{Feature squeezing (FS) \cite{xu2017feature}}: FS is a filter-based approach that applies filters to a given image and measures the distance between prediction vectors of the two images. If the distance for any compressed image exceeds a certain threshold learned from benign images, the unaltered image is considered adversarial.

\underline{Magnet \cite{meng2017magnet}:} MagNet is a reconstruction-based detector that trains denoisers on clean training data to reconstruct input samples. If the reconstruction error score exceeds a threshold learned from benign images, the detector flags an input sample as adversarial.

 \underline{Turning a weakness into a strength (TWS) \cite{hu2019new}:} TWS is a noise-based approach that identifies a given input image as adversarial if after perturbing the input with noise does not result in a significant change in softmax score. The defense also has a second evaluation criterion, which checks the number of steps required to cross the decision boundary to a random target class. The second test assumes white-box access to the model and detector and requires modification of the adversarial attack. Hence, we only use the first criteria as the detection mechanism. 

\begin{table*}[]
\centering
\caption{Adversarial Detection Performance for MNIST and CIFAR-10 models: Our Method (PASA) vs. Unsupervised Methods (FS, MagNet, U-LOO, TWS) using TPR scores.}
\label{tab:mnistcifartpr}
\resizebox{\textwidth}{!}{%
\begin{tabular}{@{}l|l|ccccc|ccccc|ccccc@{}}
\toprule
\textbf{} &
  \textbf{Performance} &
  \multicolumn{5}{c|}{\textbf{MNIST}} &
  \multicolumn{5}{c|}{\textbf{CIFAR-10 (VGG)}} &
  \multicolumn{5}{c}{\textbf{CIFAR-10 (ResNet)}} \\ \midrule
\textbf{Attack} &
  \textbf{Metric} &
  \textbf{FS} &
  \textbf{MagNet} &
  \textbf{U-LOO} &
  \textbf{TWS} &
  \textbf{PASA} &
  \textbf{FS} &
  \textbf{MagNet} &
  \textbf{U-LOO} &
  \textbf{TWS} &
  \textbf{PASA} &
  \textbf{FS} &
  \textbf{MagNet} &
  \textbf{U-LOO} &
  \textbf{TWS} &
  \textbf{PASA} \\ \midrule
\textbf{FGSM (8/255)}   & \textbf{TPR (FPR @ 0.01)} & 85.9 & 74.8 & 40.36 & 6.4  & 90.3 & 11.8 & 5.5  & 3.5  & 2.6  & 18.8 & 29.4 & 4.6  & 1    & 12.5 & 22.6  \\
                        & \textbf{TPR (FPR @ 0.05)} & 96.5 & 95.3 & 91.9  & 64   & 97   & 21.2 & 5.6  & 9.9  & 7.5  & 23.3 & 33.1 & 16.8 & 5.3  & 15.4 & 22.9  \\
                        & \textbf{TPR (FPR @ 0.1)}  & 96.8 & 98.3 & 92.3  & 86.8 & 99.8 & 26.2 & 19.6 & 16.8 & 11.7 & 31.7 & 37.4 & 37.2 & 8.3  & 17.8 & 56.8  \\ \midrule
\textbf{FGSM (16/255)}  & \textbf{TPR (FPR @ 0.01)} & 86.3 & 95.3 & 51.5  & 3.5  & 96.1 & 13.8 & 9.9  & 3.4  & 2.8  & 17   & 31.7 & 10.9 & 0.7  & 9.5  & 64.9  \\
                        & \textbf{TPR (FPR @ 0.05)} & 83   & 96.7 & 81.2  & 59.4 & 98.7 & 19.4 & 10.6 & 9.2  & 8.8  & 32.2 & 36   & 11   & 4.8  & 14.8 & 97.1  \\
                        & \textbf{TPR (FPR @ 0.1)}  & 85.4 & 99.4 & 91.3  & 93.1 & 99.9 & 27.7 & 42.6 & 15.7 & 14.2 & 32.3 & 42.7 & 43.4 & 7.6  & 18.5 & 100   \\ \midrule
\textbf{FGSM (32/255)}  & \textbf{TPR (FPR @ 0.01)} & 68.8 & 94.9 & 60.9  & 6.3  & 98.4 & 10.9 & 98.2 & 2    & 1.3  & 16.2 & 10.7 & 98.8 & 1.7  & 0.7  & 100   \\
                        & \textbf{TPR (FPR @ 0.05)} & 69.2 & 96.8 & 90.9  & 47.5 & 99.6 & 15.2 & 98.6 & 8    & 3.8  & 30.2 & 12.9 & 98.8 & 5.7  & 12   & 100   \\
                        & \textbf{TPR (FPR @ 0.1)}  & 81.3 & 99.1 & 90.11 & 88   & 99.7 & 23.5 & 98.8 & 14.6 & 7.6  & 51.8 & 18.4 & 99.1 & 9.3  & 21   & 100   \\ \midrule
\textbf{FGSM (64/255)}  & \textbf{TPR (FPR @ 0.01)} & 84.4 & 99.9 & 78.8  & 5.4  & 99.7 & 5.8  & 92.6 & 1.2  & 0.5  & 36.5 & 16.8 & 100  & 0.2  & 0.1  & 100   \\
                        & \textbf{TPR (FPR @ 0.05)} & 87.9 & 99.9 & 88.9  & 47.3 & 99.9 & 8.4  & 93.6 & 4.9  & 0.8  & 76.7 & 24.3 & 100  & 0.2  & 20   & 100   \\
                        & \textbf{TPR (FPR @ 0.1)}  & 91.8 & 99.9 & 90.5  & 83.8 & 100  & 13.4 & 94.6 & 9.8  & 1.6  & 91.8 & 32.9 & 100  & 4.2  & 78.8 & 100   \\ \midrule
\textbf{PGD (8/255)}    & \textbf{TPR (FPR @ 0.01)} & 78.9 & 100  & 100   & 13.9 & 100  & 5.8  & 4.5  & 1.2  & 3    & 53.7 & 1.7  & 4.3  & 2.8  & 0    & 12.8  \\
                        & \textbf{TPR (FPR @ 0.05)} & 79.9 & 100  & 100   & 49.6 & 100  & 6.4  & 4.8  & 4.9  & 6    & 55   & 1.8  & 4.3  & 10.8 & 0    & 25.2  \\
                        & \textbf{TPR (FPR @ 0.1)}  & 99.3 & 100  & 100   & 90.2 & 100  & 7.5  & 17.4 & 12.5 & 8    & 55.3 & 3.1  & 15   & 14.9 & 0    & 52    \\ \midrule
\textbf{PGD (16/255)}   & \textbf{TPR (FPR @ 0.01)} & 72.5 & 100  & 100   & 1.6  & 100  & 4    & 6.6  & 2    & 5    & 66.5 & 0.6  & 6.1  & 3.1  & 0    & 40.9  \\
                        & \textbf{TPR (FPR @ 0.05)} & 77.8 & 100  & 100   & 13.6 & 100  & 4.2  & 7    & 6.2  & 9    & 67.8 & 1.3  & 6.2  & 11.1 & 0    & 61    \\
                        & \textbf{TPR (FPR @ 0.1)}  & 85.6 & 100  & 100   & 54.9 & 100  & 4.3  & 24.1 & 11.9 & 12   & 68.2 & 2    & 22.9 & 16.1 & 0    & 86    \\ \midrule
\textbf{PGD (32/255)}   & \textbf{TPR (FPR @ 0.01)} & 44.3 & 100  & 100   & 1.3  & 100  & 8.6  & 34.8 & 2.5  & 0.5  & 72.7 & 0.2  & 30.5 & 5.8  & 0    & 70.6  \\
                        & \textbf{TPR (FPR @ 0.05)} & 48.9 & 100  & 100   & 5.8  & 100  & 10.1 & 37.5 & 7.4  & 3.5  & 77.1 & 0.2  & 31.6 & 16.4 & 0    & 95.2  \\
                        & \textbf{TPR (FPR @ 0.1)}  & 53.6 & 100  & 100   & 9    & 100  & 12.8 & 100  & 14.3 & 5    & 78.2 & 0.3  & 100  & 22.7 & 0    & 99.8  \\ \midrule
\textbf{PGD (64/255)}   & \textbf{TPR (FPR @ 0.01)} & 12.6 & 100  & 100   & 0.1  & 100  & 8    & 100  & 1.6  & 5    & 95.4 & 0.1  & 100  & 7.2  & 0    & 100   \\
                        & \textbf{TPR (FPR @ 0.05)} & 17.9 & 100  & 100   & 0.9  & 100  & 8.9  & 100  & 4.4  & 6    & 96.2 & 0.1  & 100  & 17.9 & 0    & 100   \\
                        & \textbf{TPR (FPR @ 0.1)}  & 24.6 & 100  & 100   & 1.5  & 100  & 11.3 & 100  & 12   & 8    & 96.6 & 0.8  & 100  & 23.3 & 0    & 100   \\ \midrule
\textbf{BIM (8/255)}    & \textbf{TPR (FPR @ 0.01)} & 72   & 38.9 & 4.4   & 4    & 3    & 7.8  & 5.1  & 2.1  & 2.1  & 34.8 & 2.8  & 6    & 1.9  & 0    & 3.2   \\
                        & \textbf{TPR (FPR @ 0.05)} & 82.5 & 45.6 & 6.9   & 52   & 8.2  & 11.5 & 5.4  & 6.9  & 5.1  & 37.1 & 3.2  & 16.8 & 7.6  & 0    & 10.9  \\
                        & \textbf{TPR (FPR @ 0.1)}  & 92.9 & 56.7 & 12.5  & 96.4 & 16.1 & 18.7 & 15.9 & 13.5 & 7.4  & 38   & 3.6  & 28.9 & 12.3 & 0    & 37.7  \\ \midrule
\textbf{BIM (16/255)}   & \textbf{TPR (FPR @ 0.01)} & 76.6 & 65.5 & 9.8   & 2.8  & 2.7  & 7.8  & 4.6  & 1.2  & 3.2  & 49.4 & 0.9  & 5.9  & 2.3  & 0    & 11.5  \\
                        & \textbf{TPR (FPR @ 0.05)} & 86.8 & 88.5 & 37.3  & 39.5 & 7.6  & 13   & 4.7  & 6.5  & 6.5  & 50.2 & 0.9  & 16.1 & 7.6  & 0    & 26.5  \\
                        & \textbf{TPR (FPR @ 0.1)}  & 97   & 91.1 & 50.5  & 91.5 & 14   & 16.1 & 16.2 & 13.2 & 11.2 & 50.8 & 1.1  & 32.2 & 11.8 & 0    & 54.3  \\ \midrule
\textbf{BIM (32/255)}   & \textbf{TPR (FPR @ 0.01)} & 75.4 & 75.5 & 26.3  & 0.8  & 3.7  & 9.1  & 5.4  & 2.1  & 2.8  & 54.3 & 0.4  & 6.4  & 1.2  & 0    & 23.6  \\
                        & \textbf{TPR (FPR @ 0.05)} & 85.7 & 99.4 & 65    & 28.2 & 8.2  & 12.5 & 5.6  & 7.4  & 7.4  & 55.7 & 0.4  & 22.6 & 6.6  & 0    & 51.5  \\
                        & \textbf{TPR (FPR @ 0.1)}  & 96.1 & 99.5 & 75.1  & 74.6 & 14   & 18.7 & 24.5 & 14.2 & 12   & 56.6 & 0.6  & 49.1 & 10.3 & 0    & 83.3  \\ \midrule
\textbf{BIM (64/255)}   & \textbf{TPR (FPR @ 0.01)} & 63.9 & 100  & 53.6  & 0.3  & 4    & 15.1 & 12.6 & 1.9  & 3    & 37.2 & 0.4  & 12.8 & 1.2  & 0    & 958.2 \\
                        & \textbf{TPR (FPR @ 0.05)} & 74.3 & 100  & 86.8  & 9.3  & 8.5  & 19.4 & 14.5 & 8.2  & 8.1  & 40.3 & 0.6  & 88   & 6.1  & 0    & 94.4  \\
                        & \textbf{TPR (FPR @ 0.1)}  & 84.4 & 100  & 92.3  & 44.1 & 14.3 & 16.6 & 86.4 & 14.4 & 11.2 & 42.1 & 0.6  & 100  & 9.2  & 0    & 99.9  \\ \midrule
\textbf{Auto-PGD (0.15)} & \textbf{TPR (FPR @ 0.01)} & 90.7 & 100  & 83    & 0.15 & 99.2 & 0    & 82.5 & 2    & 0    & 98.2 & 0.4  & 51   & 8.7  & 0    & 85.4  \\
                        & \textbf{TPR (FPR @ 0.05)} & 91   & 100  & 96    & 3.4  & 99.3 & 0    & 83.6 & 7.8  & 0    & 98.8 & 0.4  & 51.8 & 23.4 & 0    & 98.7  \\
                        & \textbf{TPR (FPR @ 0.1)}  & 91.6 & 100  & 97    & 17.3 & 99.6 & 0    & 84.8 & 15.3 & 0    & 98.8 & 1.2  & 100  & 29.7 & 0    & 99.8  \\ \midrule
\textbf{CW (0.15)}      & \textbf{TPR (FPR @ 0.01)} & 85.3 & 86.3 & 29.1  & 3.9  & 2.1  & 14.6 & 26.4 & 3.5  & 2.8  & 6.8  & 38.7 & 57.2 & 2.3  & 7.3  & 82.2  \\
                        & \textbf{TPR (FPR @ 0.05)} & 87.9 & 96.6 & 70.4  & 62.8 & 3.4  & 21.5 & 28.6 & 9.3  & 7.4  & 20.2 & 45.4 & 57.5 & 7.3  & 13   & 97.8  \\
                        & \textbf{TPR (FPR @ 0.1)}  & 93.1 & 96.8 & 80.6  & 98.1 & 13.8 & 29.2 & 90.4 & 16.8 & 13.3 & 42.1 & 52.2 & 87  & 9.8  & 18.9 & 99.7  \\ \bottomrule
\end{tabular}%
}
\end{table*}

 \underline{ML-LOO \cite{yang2020ml}}: ML-LOO is a feature-attribution-based defense that detects adversarial examples using statistical measures of attribution vector. The authors compute the inter-quartile range (IQR) of feature attribution of benign and adversarial images for distinguishing benign images from adversarial counterparts. While the paper also proposes a supervised detector by extracting statistics from multiple hidden layers, we implement unsupervised detection, U-LOO, for a fair comparison. The authors evaluate their results using LOO \cite{li2016understanding} and IG \cite{sundararajan2017axiomatic}. We stick to IG since our detection also uses the same method.

\subsubsection{{Performance evaluation}}

For each dataset, we randomly sample 1000 benign samples from the test set, which are correctly classified by the model, and generate 1000 adversarial samples for each type of attack for evaluation. This is repeated 10 times to account for randomness associated with the sampling. During test time, \textit{we assume no previous knowledge of the attack mechanism}. Given an input sample, PASA only computes two noise-probed metrics, prediction sensitivity, and attribution sensitivity, and compares them with the threshold learned during training. If either of the metrics satisfies the threshold, the sample is classified as benign, else adversarial.

\paragraph{\textbf{Metrics}} We assess the detector performance using in the following criteria: \textit{a) True Positive Rate (TPR):} TPR is computed as the ratio of the total number of correctly identified adversarial samples to the overall number of adversarial samples. In unsupervised detectors, the decision threshold is learned from benign samples while maintaining a fixed false positive rate (FPR) on the validation set. We then use this threshold on the test set and compute the TPR. We report the TPR of detectors using thresholds calculated for 1\%, 5\%, and 10\% FPR. \textit{b) Area Under the Receiver Operating Characteristic Curve (AUC):} AUC is a threshold-independent measure of a detector performance which is widely used as a standard in comparison between different methods \cite{davis2006relationship}.

\begin{table*}[]
\centering
\caption{Adversarial Detection Performance for CIFAR-100 and ImageNet models: Our Method (PASA) vs. Unsupervised Methods (FS, MagNet, U-LOO, TWS) using AUC scores.}
\label{tab:cifarimagenetresult}
\resizebox{\textwidth}{!}{%
\begin{tabular}{@{}lclllll|lllll|lllll@{}}
\toprule
\textbf{Attack} &
  \multicolumn{1}{l}{\textbf{}} &
  \multicolumn{5}{c|}{\textbf{CIFAR-100}} &
  \multicolumn{5}{c|}{\textbf{ImageNet (MobileNet)}} &
  \multicolumn{5}{c}{\textbf{ImageNet (ResNet)}} \\ \midrule
\textbf{} &
  \textbf{Strength} &
  \multicolumn{1}{c}{\textbf{FS}} &
  \multicolumn{1}{c}{\textbf{MagNet}} &
  \multicolumn{1}{c}{\textbf{U-LOO}} &
  \multicolumn{1}{c}{\textbf{TWS}} &
  \multicolumn{1}{c|}{\textbf{PASA}} &
  \multicolumn{1}{c}{\textbf{FS}} &
  \multicolumn{1}{c}{\textbf{MagNet}} &
  \multicolumn{1}{c}{\textbf{U-LOO}} &
  \multicolumn{1}{c}{\textbf{TWS}} &
  \multicolumn{1}{c|}{\textbf{PASA}} &
  \multicolumn{1}{c}{\textbf{FS}} &
  \multicolumn{1}{c}{\textbf{MagNet}} &
  \multicolumn{1}{c}{\textbf{U-LOO}} &
  \multicolumn{1}{c}{\textbf{TWS}} &
  \multicolumn{1}{c}{\textbf{PASA}} \\ \midrule
\textbf{FGSM} &
  \textbf{8/255} &
  0.68$\pm$0.02 &
  0.60$\pm$0.02 &
  0.62$\pm$0.03 &
  0.34$\pm$0.02 &
  \textbf{0.81$\pm$0.01} &
  0.60$\pm$0.01 &
  0.51$\pm$0.03 &
  0.60$\pm$0.01 &
  0.50$\pm$0.02 &
  \textbf{0.81$\pm$0.01} &
  0.65$\pm$0.02 &
  0.50$\pm$0.01 &
  0.62$\pm$0.01 &
  0.64$\pm$0.01 &
  \textbf{0.65$\pm$0.01} \\
\textbf{} &
  \textbf{16/255} &
  0.62$\pm$0.04 &
  0.78$\pm$0.03 &
  0.67$\pm$0.02 &
  0.31$\pm$0.01 &
  \textbf{0.96$\pm$0.01} &
  0.62$\pm$0.01 &
  0.51$\pm$0.03 &
  0.68$\pm$0.01 &
  0.50$\pm$0.03 &
  \textbf{0.91$\pm$0.02} &
  0.69$\pm$0.01 &
  0.52$\pm$0.01 &
  0.62$\pm$0.01 &
  0.57$\pm$0.01 &
  \textbf{0.75$\pm$0.01} \\
\textbf{} &
  \textbf{32/255} &
  0.64$\pm$0.03 &
  0.95$\pm$0.02 &
  0.67$\pm$0.02 &
  0.29$\pm$0.01 &
  \textbf{0.97$\pm$0.02} &
  0.65$\pm$0.01 &
  0.54$\pm$0.02 &
  0.69$\pm$0.01 &
  0.50$\pm$0.04 &
  \textbf{0.96$\pm$0.01} &
  0.75$\pm$0.02 &
  0.55$\pm$0.01 &
  0.60$\pm$0.01 &
  0.52$\pm$0.01 &
  \textbf{0.86$\pm$0.01} \\
\textbf{} &
  \textbf{64/255} &
  0.68$\pm$0.02 &
  0.96$\pm$0.02 &
  0.67$\pm$0.03 &
  0.24$\pm$0.02 &
  \textbf{0.97$\pm$0.02} &
  0.68$\pm$0.01 &
  0.63$\pm$0.03 &
  0.68$\pm$0.01 &
  0.49$\pm$0.03 &
  \textbf{0.98$\pm$0.01} &
  0.81$\pm$0.01 &
  0.65$\pm$0.01 &
  0.60$\pm$0.01 &
  0.45$\pm$0.01 &
  \textbf{0.95$\pm$0.01} \\  \midrule
\textbf{PGD} &
  \textbf{8/255} &
  \textbf{0.67$\pm$0.03} &
  0.56$\pm$0.03 &
  0.63$\pm$0.03 &
  0.61$\pm$0.01 &
  0.60$\pm$0.02 &
  0.25$\pm$0.02 &
  0.51$\pm$0.01 &
  0.59$\pm$0.01 &
  0.52$\pm$0.01 &
  \textbf{0.98$\pm$0.02} &
  0.29$\pm$0.01 &
  0.51$\pm$0.01 &
  0.59$\pm$0.01 &
  0.57$\pm$0.02 &
  \textbf{0.97$\pm$0.01} \\
 &
  \textbf{16/255} &
  0.62$\pm$0.02 &
  0.66$\pm$0.04 &
  0.68$\pm$0.03 &
  0.59$\pm$0.01 &
  \textbf{0.68$\pm$0.02} &
  0.19$\pm$0.03 &
  0.50$\pm$0.01 &
  0.59$\pm$0.01 &
  0.51$\pm$0.02 &
  \textbf{0.99$\pm$0.02} &
  0.18$\pm$0.01 &
  0.52$\pm$0.02 &
  0.63$\pm$0.02 &
  0.27$\pm$0.02 &
  \textbf{0.98$\pm$0.01} \\
 &
  \textbf{32/255} &
  0.74$\pm$0.03 &
  0.85$\pm$0.05 &
  0.72$\pm$0.03 &
  0.48$\pm$0.02 &
  \textbf{0.86$\pm$0.04} &
  0.16$\pm$0.02 &
  0.52$\pm$0.01 &
  0.57$\pm$0.01 &
  0.51$\pm$0.01 &
  \textbf{0.98$\pm$0.02} &
  0.11$\pm$0.02 &
  0.52$\pm$0.01 &
  0.75$\pm$0.02 &
  0.05$\pm$0.00 &
  \textbf{0.97$\pm$0.01} \\
 &
  \textbf{64/255} &
  0.69$\pm$0.02 &
  0.95$\pm$0.03 &
  0.73$\pm$0.03 &
  0.08$\pm$0.01 &
  \textbf{0.95$\pm$0.02} &
  0.17$\pm$0.02 &
  0.57$\pm$0.01 &
  0.59$\pm$0.01 &
  0.50$\pm$0.01 &
  \textbf{0.98$\pm$0.02} &
  0.11$\pm$0.02 &
  0.57$\pm$0.01 &
  0.74$\pm$0.03 &
  0.02$\pm$0.00 &
  \textbf{0.98$\pm$0.01} \\  \midrule
\textbf{BIM} &
  \textbf{8/255} &
  0.55$\pm$0.02 &
  0.51$\pm$0.01 &
  0.57$\pm$0.02 &
  0.59$\pm$0.02 &
  \textbf{0.60$\pm$0.01} &
  0.42$\pm$0.03 &
  0.02$\pm$0.00 &
  0.19$\pm$0.01 &
  0.46$\pm$0.01 &
  \textbf{0.51$\pm$0.02} &
  0.50$\pm$0.03 &
  0.04$\pm$0.00 &
  0.15$\pm$0.01 &
 \textbf{0.81$\pm$0.01} &
  0.37$\pm$0.01 \\
 &
  \textbf{16/255} &
  0.50$\pm$0.01 &
  0.55$\pm$0.02 &
  0.58$\pm$0.02 &
  0.56$\pm$0.04 &
  \textbf{0.61$\pm$0.03} &
  0.32$\pm$0.02 &
  0.03$\pm$0.00 &
  0.15$\pm$0.02 &
  0.51$\pm$0.03 &
  \textbf{0.63$\pm$0.01} &
  0.31$\pm$0.02 &
  0.04$\pm$0.00 &
  0.17$\pm$0.01 &
  \textbf{0.71$\pm$0.01} &
  0.47$\pm$0.01 \\
 &
  \textbf{32/255} &
  0.57$\pm$0.01 &
  0.65$\pm$0.03 &
  0.62$\pm$0.03 &
  0.42$\pm$0.02 &
  \textbf{0.70$\pm$0.02} &
  0.25$\pm$0.01 &
  0.02$\pm$0.00 &
  0.16$\pm$0.01 &
  0.51$\pm$0.02 &
  \textbf{0.77$\pm$0.01} &
  0.20$\pm$0.02 &
  0.04$\pm$0.00 &
  0.17$\pm$0.01 &
  0.58$\pm$0.02 &
  \textbf{0.62$\pm$0.01} \\
 &
  \textbf{64/255} &
  0.54$\pm$0.01 &
  0.83$\pm$0.02 &
  0.61$\pm$0.02 &
  0.33$\pm$0.03 &
  \textbf{0.84$\pm$0.03} &
  0.21$\pm$0.01 &
  0.02$\pm$0.00 &
  0.18$\pm$0.01 &
  0.50$\pm$0.02 &
  \textbf{0.83$\pm$0.03} &
  0.15$\pm$0.02 &
  0.04$\pm$0.00 &
  0.18$\pm$0.02 &
  0.29$\pm$0.03 &
  \textbf{0.62$\pm$0.01} \\  \midrule
\textbf{Auto-PGD} &
  \textbf{0.15} &
  0.32$\pm$0.01 &
  0.91$\pm$0.02 &
  0.64$\pm$0.04 &
  0.30$\pm$0.01 &
  \textbf{0.98$\pm$0.02} &
  0.19$\pm$0.04 &
  0.37$\pm$0.01 &
  0.59$\pm$0.01 &
  0.17$\pm$0.02 &
  \textbf{0.97$\pm$0.02} &
  0.14$\pm$0.01 &
  0.38$\pm$0.01 &
  0.66$\pm$0.01 &
  0.10$\pm$0.00 &
  \textbf{0.96$\pm$0.01} \\  \midrule
\textbf{CW} &
  \textbf{0.15} &
  0.58$\pm$0.02 &
  0.91$\pm$0.01 &
  0.68$\pm$0.03 &
  0.22$\pm$0.01 &
  \textbf{0.92$\pm$0.03} &
  0.58$\pm$0.01 &
  0.03$\pm$0.00 &
  0.11$\pm$0.02 &
  0.51$\pm$0.01 &
  \textbf{0.87$\pm$0.01} &
  0.58$\pm$0.01 &
  0.04$\pm$0.00 &
  0.15$\pm$0.01 &
  0.68$\pm$0.02 &
  \textbf{0.95$\pm$0.01} \\  \midrule
\textbf{Average} &
  \multicolumn{1}{l}{} &
  0.60$\pm$0.10 &
  0.77$\pm$0.16 &
  0.65$\pm$0.05 &
  0.38$\pm$0.16 &
  \textbf{0.81$\pm$0.15} &
  0.38$\pm$0.20 &
  0.34$\pm$0.24 &
  0.45$\pm$0.22 &
  0.48$\pm$0.09 &
  \textbf{0.87$\pm$0.15} &
  0.39$\pm$0.25 &
  0.35$\pm$0.24 &
  0.47$\pm$0.23 &
  0.45$\pm$0.25 &
  \textbf{0.79$\pm$0.20} \\ \bottomrule
\end{tabular}%
}
\end{table*}

\section{{Results and Analysis}}\label{sec:resultandanalysis}
We first discuss the results of adversarial detection on image classifiers. We discuss the performance of detectors on the security dataset in Section \ref{sec:securitydataset}.

\subsection{Adversarial Detection Performance}
\paragraph{\textbf{CIFAR-10}} PASA outperforms all baseline methods in CIFAR-10 (ResNet) model. For example, as observed in Table \ref{tab:mnistcifarresult}, PASA obtains an AUC of 0.98$\pm$0.01 for detecting CW attack on CIFAR-10 ResNet. The next best detector is MagNet with 0.93$\pm$0.01 AUC. On CIFAR-10 (VGG) model, PASA obtains an AUC of 0.82$\pm$0.01 for detecting CW attack. MagNet is the the next best detector with 0.71$\pm$0.03 AUC. Other methods (e.g., FS, and TWS) show more variability in CIFAR-10 models, with lower AUC scores (less than 0.5) in some instances, suggesting that the thresholds learned from benign images were suitable for detecting specific attacks only. Thus such solutions become impractical for attack-agnostic detection since they require a threshold change depending on the type of attack.

We also observe that the MagNet performance remains competitive on both CIFAR-10 models, however, the performance of other detection methods degrades significantly. For example, on average, the U-LOO method obtains an AUC of 0.90$\pm$0.16 on MNIST, whereas the average AUC reduces to 0.51$\pm$0.01 on the CIFAR-10 VGG model and 0.61$\pm$0.08 on the ResNet model. A similar performance drop can be observed with TWS and FS. In Table \ref{tab:mnistcifartpr}, we notice that PASA obtains high TPRs at low FPRs consistently. For example, PASA obtains a TPR of 82.2\% at 1\% FPR on detecting CW attack for the CIFAR-10 (ResNet) model. The next best detector is MagNet with 57.2\% TPR. While MagNet seems to obtain high TPRs, especially on the VGG16 model, it comes at the cost of high FPRs, discussed in Section \ref{sec:falsepositiverate}.

\paragraph{\textbf{CIFAR-100}} As observed in Table \ref{tab:cifarimagenetresult}, PASA consistently outperforms the baseline methods on CIFAR-100 with noticeable performance improvement as the strength of adversarial perturbation increases. While the performance of detectors like TWS decreases with an increase in adversarial perturbation, PASA achieves an increment in its detection performance. This is because as perturbation increases, the discrepancy between the attribution maps of benign and adversarial images increases, which helps PASA detect the inconsistency. For instance, PASA obtains an AUC of 0.92$\pm$0.03 on detecting CW attacks. The next best detector is MagNet, with 0.91$\pm$0.01 AUC. TWS, also a noise-based approach, obtains an AUC of 0.59$\pm$0.02 on detecting $\epsilon=8/255$ BIM attack. The AUC reduces to 0.33$\pm$0.03 at $\epsilon=64/255$. PASA, on the other hand, improves from an AUC of 0.60$\pm$0.01 to 0.84$\pm$0.03 on detecting $\epsilon=8/255$ and $\epsilon=64/255$ BIM attacks. Averaged across all attacks, PASA obtains an AUC of 0.81$\pm$0.15, with the next best detector, MagNet, obtaining 0.77$\pm$0.16 AUC. In Table \ref{tab:cifarimagenettpr}, we can observe that PASA obtains the highest TPRs at the lowest FPRs settings consistently. For example, PASA obtains a TPR of 34.5\% on detecting CW attacks at 1\% FPR. The next best detector is FS with 26.7\% TPR.

\paragraph{\textbf{ImageNet}} Table \ref{tab:cifarimagenetresult} demonstrates that PASA consistently outperforms the baseline methods in detecting attacks on both ImageNet models. For instance, when detecting an $\epsilon=8/255$ PGD attack on ImageNet (MobileNet) and ImageNet (ResNet), PASA scores an AUC of 0.98$\pm$0.02 and 0.97$\pm$0.01 respectively, outperforming all baselines by a significant margin. In the ImageNet-ResNet model, while PASA obtains an AUC of 0.95$\pm$0.01 for the CW attack, the next best detector only has an AUC of 0.66$\pm$0.01. Baseline method (TWS) performance is slightly better than PASA in detecting two BIM attacks (8/255 and 16/255) on ImageNet (ResNet). However, as the attack strength increases, our method surpasses the performance of TWS. MagNet and U-LOO have very low AUC scores on attacks like CW for ImageNet, which means that the threshold learned from benign images was only suitable for detecting FGSM and PGD attacks. This suggests that the detection criteria used in those methods may not be effective against different types of attacks without knowing the attack types beforehand, which is impractical. From Table \ref{tab:cifarimagenettpr}, we can observe that PASA obtains the highest TPRs at low FPRs across different attacks in both ImageNet models.

\paragraph{\textbf{MNIST}} All unsupervised detectors have overall competitive performance in detecting adversarial attacks on MNIST, with MagNet consistently obtaining high AUC and TPR scores. However, PASA has a drop in performance, especially in detecting BIM and CW attacks. This could be attributed to the lower resolution of MNIST images. Lower resolution (28x28) implies less visual information for the feature attribution method, compared with CIFAR-10 (32x32x3) and ImageNet (224x224x3). It limits the granularity at which IG attributes importance to individual features, resulting in a small number of attributions and lower sensitivity to noise.

\subsubsection{Analysis} PASA leverages the discrepancy between benign and adversarial samples in model logits and class-specific feature attribution for detecting adversarially perturbed samples. This discrepancy can be measured by injecting noise into a given input sample and measuring the change in both logits and attribution maps caused by noise. Previous studies  \cite{hu2019new} have demonstrated that the model response of benign and adversarial inputs to noise differs because neural networks are trained only with benign inputs. In this work, we also demonstrate that the sensitivity of Integrated Gradients (IG) attribution is linked to the sensitivity of the model (Eqn. \ref{eqn:motivationatr}). However, since IG assigns importance to each feature, the level of granularity in importance attribution depends on the number of features. Therefore, based on these considerations, we combined these two inconsistency measures for the detection of adversarially perturbed samples and we developed PASA accordingly to account for \textit{a) the sensitivity of the trained model to noise}, and \textit{b) the granularity of IG attribution}. 

In our experiments, we followed a standard classification pipeline to achieve high performance on the test set, using standard hyperparameters or pretrained models (training details are discussed in Appendix \ref{appendix:models}). However, different deep learning models learn varying levels of feature abstraction from a given dataset due to differences in depth, connections, and overall structure. For instance, ResNet, with its residual connections, can capture more intricate features compared to simpler networks like VGG or LeNet. Our experimental results indicate that PASA performs notably better with deeper networks, as demonstrated by the results obtained from ResNet models trained on CIFAR-10 and ImageNet (refer to Tables \ref{tab:mnistcifarresult} and \ref{tab:cifarimagenetresult}). These findings suggest that increased network depth, as observed in ResNet, enables the model to extract complex patterns from the dataset, resulting in higher sensitivity to noise—a quality utilized by PASA for detecting adversarial samples. Additionally, the level of granularity in IG attribution depends on the number of features present in the dataset. Consequently, PASA exhibits a notable decrease in detecting certain attacks (e.g., BIM, CW) on MNIST, as the lower resolution of MNIST leads to smaller norms of IG attribution. However, it consistently obtains high detection performance on varying attacks on CIFAR-10, CIFAR-100, and ImageNet.

\begin{figure}[]
    \centering
\includegraphics[width=.28\textwidth]{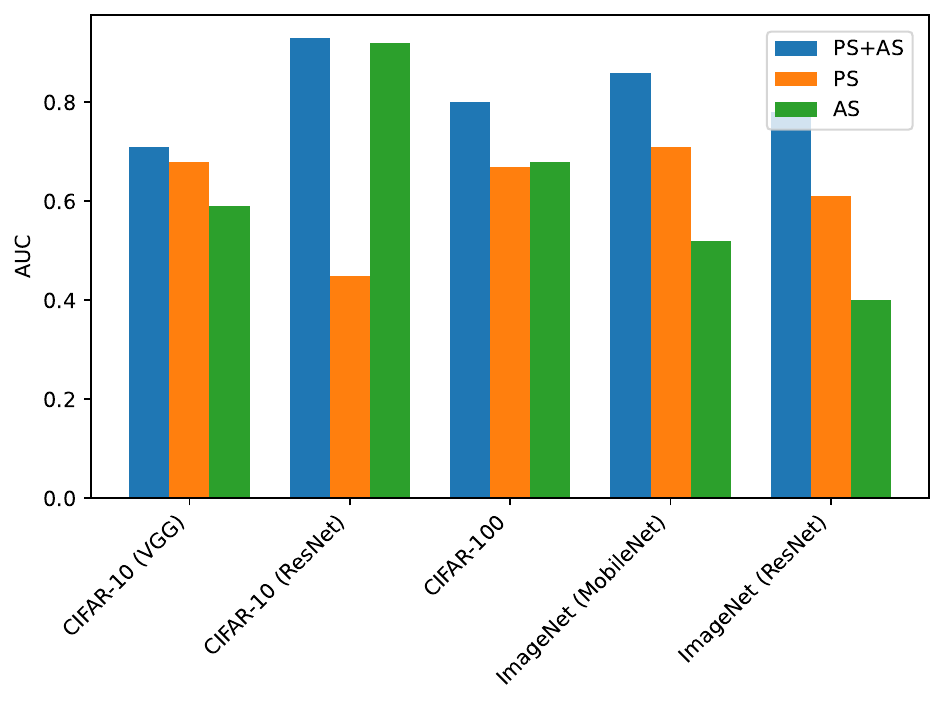}
    \caption{Comparing adversarial detection performance of PASA with its components, PS: Prediction Sensitivity, AS: Attribution Sensitivity, and PS+AS: PASA.}
    \label{fig:ablationstudy}
\end{figure}

\begin{table*}[]
\centering
\caption{Adversarial Detection Performance for CIFAR-100 and ImageNet models: Our Method (PASA) vs. Unsupervised Methods (FS, MagNet, U-LOO, TWS) using TPR scores.}
\label{tab:cifarimagenettpr}
\resizebox{\textwidth}{!}{%
\begin{tabular}{@{}l|l|ccccc|ccccc|ccccc@{}}
\toprule
\textbf{} &
  \textbf{Performance} &
  \multicolumn{5}{c|}{\textbf{CIFAR-100}} &
  \multicolumn{5}{c|}{\textbf{ImageNet (MobileNet)}} &
  \multicolumn{5}{c}{\textbf{ImageNet (ResNet)}} \\ \midrule
\textbf{Attack} &
  \textbf{Metric} &
  \textbf{FS} &
  \textbf{MagNet} &
  \textbf{U-LOO} &
  \textbf{TWS} &
  \textbf{PASA} &
  \textbf{FS} &
  \textbf{MagNet} &
  \textbf{U-LOO} &
  \textbf{TWS} &
  \textbf{PASA} &
  \textbf{FS} &
  \textbf{MagNet} &
  \textbf{U-LOO} &
  \textbf{TWS} &
  \textbf{PASA} \\ \midrule
\textbf{FGSM (8/255)}   & \textbf{TPR (FPR @ 0.01)} & 6.2  & 4.1  & 12.3 & 5.5 & 10.3 & 2    & 6.1  & 2.1  & 0.5  & 11.2 & 7.1  & 4.3 & 2    & 2.3  & 3.9   \\
                        & \textbf{TPR (FPR @ 0.05)} & 25.1 & 4.4  & 15.7 & 4.6 & 35.9 & 4.4  & 6.9  & 13.4 & 6.9  & 29.5 & 11.3 & 5   & 7.1  & 7.4  & 15.1  \\
                        & \textbf{TPR (FPR @ 0.1)}  & 36.1 & 15.1 & 17.8 & 5.8 & 54.8 & 11   & 16.3 & 21.8 & 16.1 & 49.8 & 15.4 & 5.2 & 18.1 & 32.5 & 27.6  \\ \midrule
\textbf{FGSM (16/255)}  & \textbf{TPR (FPR @ 0.01)} & 7.5  & 4.5  & 17.3 & 5.9 & 61.3 & 2.2  & 4.6  & 2.1  & 1.3  & 29.5 & 8.8  & 5.3 & 1.2  & 0.3  & 2.6   \\
                        & \textbf{TPR (FPR @ 0.05)} & 29.3 & 4.7  & 28.7 & 4.6 & 86.3 & 4.5  & 5.3  & 15   & 6.5  & 56.7 & 12.8 & 6.1 & 8.2  & 1.7  & 19    \\
                        & \textbf{TPR (FPR @ 0.1)}  & 47.3 & 27   & 37.3 & 5.1 & 94   & 15.9 & 14.3 & 25   & 13.8 & 79   & 15.9 & 6.3 & 21.5 & 12.4 & 33.3  \\ \midrule
\textbf{FGSM (32/255)}  & \textbf{TPR (FPR @ 0.01)} & 25.5 & 82.5 & 17.4 & 1.2 & 96.6 & 3.5  & 5.1  & 2.1  & 0.5  & 74.6 & 10.9 & 4.8 & 1.3  & 0.1  & 14.7  \\
                        & \textbf{TPR (FPR @ 0.05)} & 33.4 & 84.1 & 15.3 & 7.8 & 99.7 & 4.2  & 6.1  & 18.8 & 6.8  & 93.2 & 16.6 & 5.3 & 7    & 0.1  & 45.1  \\
                        & \textbf{TPR (FPR @ 0.1)}  & 60.5 & 100  & 33.3 & 6.9 & 99.9 & 35.9 & 16.4 & 29.3 & 13.8 & 97.7 & 23.8 & 6.2 & 15.8 & 3.5  & 63.4  \\ \midrule
\textbf{FGSM (64/255)}  & \textbf{TPR (FPR @ 0.01)} & 73.1 & 100  & 7.4  & 3   & 100  & 2.5  & 8.1  & 2.2  & 1.5  & 96.8 & 20.3 & 5.5 & 0.8  & 0    & 55.1  \\
                        & \textbf{TPR (FPR @ 0.05)} & 74.3 & 100  & 17.6 & 4.9 & 100  & 11.4 & 9.2  & 15.2 & 6.6  & 99.6 & 2.2  & 7.4 & 5.1  & 0    & 84.4  \\
                        & \textbf{TPR (FPR @ 0.1)}  & 75.3 & 100  & 23.7 & 5.4 & 100  & 36.6 & 21.1 & 25.6 & 11.8 & 100  & 37.2 & 8.3 & 15.3 & 0.1  & 93.36 \\ \midrule
\textbf{PGD (8/255)}    & \textbf{TPR (FPR @ 0.01)} & 22.4 & 3.8  & 15.9 & 6.3 & 3.1  & 6.2  & 6.1  & 1.6  & 0.7  & 98.8 & 5.6  & 4.2 & 5.4  & 22.2 & 99.1  \\
                        & \textbf{TPR (FPR @ 0.05)} & 27.8 & 3.9  & 24.9 & 7.2 & 23.3 & 7.1  & 7.1  & 10.8 & 5.5  & 99.1 & 6.6  & 4.7 & 15.3 & 27   & 99.1  \\
                        & \textbf{TPR (FPR @ 0.1)}  & 27.8 & 15.2 & 31.2 & 7.4 & 35.4 & 7.4  & 16.1 & 17.9 & 11.2 & 99.4 & 7.5  & 4.8 & 31.7 & 39.3 & 99.2  \\ \midrule
\textbf{PGD (16/255)}   & \textbf{TPR (FPR @ 0.01)} & 15.4 & 5.4  & 17.9 & 4.6 & 6.6  & 4.1  & 6.2  & 1.4  & 1.1  & 100  & 2.9  & 5.2 & 11.1 & 5.5  & 99.9  \\
                        & \textbf{TPR (FPR @ 0.05)} & 16.3 & 5.5  & 28.9 & 5.8 & 29.4 & 4.8  & 7    & 13   & 4.8  & 100  & 3    & 6   & 24.8 & 6    & 99.9  \\
                        & \textbf{TPR (FPR @ 0.1)}  & 17.2 & 18.7 & 35.6 & 5.4 & 45.5 & 5.1  & 14.5 & 20.6 & 12.4 & 100  & 3.7  & 6.2 & 44   & 10   & 99.9  \\ \midrule
\textbf{PGD (32/255)}   & \textbf{TPR (FPR @ 0.01)} & 8.5  & 9.9  & 15.5 & 6.4 & 14.4 & 4.4  & 7.7  & 12.1 & 2    & 100  & 1.9  & 5   & 5    & 0.4  & 100   \\
                        & \textbf{TPR (FPR @ 0.05)} & 9.2  & 10.1 & 30.7 & 7.9 & 46.1 & 5.8  & 9.1  & 11.4 & 5.98 & 100  & 2    & 6.1 & 11.3 & 0.5  & 100   \\
                        & \textbf{TPR (FPR @ 0.1)}  & 10.6 & 74.1 & 37.2 & 9.5 & 57.9 & 7.4  & 17.9 & 21.8 & 12.4 & 100  & 2.8  & 6.5 & 30   & 0.8  & 100   \\ \midrule
\textbf{PGD (64/255)}   & \textbf{TPR (FPR @ 0.01)} & 4.4  & 100  & 19.9 & 0   & 54.5 & 4.2  & 6.7  & 1.7  & 2.1  & 100  & 1.2  & 6   & 55.3 & 0    & 100   \\
                        & \textbf{TPR (FPR @ 0.05)} & 4.5  & 100  & 30.8 & 0.1 & 85.4 & 5    & 7.9  & 11.3 & 6    & 100  & 1.4  & 6.8 & 14.7 & 0    & 100   \\
                        & \textbf{TPR (FPR @ 0.1)}  & 4.5  & 100  & 38.2 & 0.1 & 96.1 & 7.8  & 18.2 & 21.4 & 14.3 & 100  & 2.7  & 6.7 & 31   & 0    & 100   \\ \midrule
\textbf{BIM (8/255)}    & \textbf{TPR (FPR @ 0.01)} & 12.3 & 4.5  & 12   & 6.7 & 4.3  & 10.1 & 0    & 0.1  & 0.5  & 27.1 & 11.5 & 0   & 61.5 & 49.1 & 14.7  \\
                        & \textbf{TPR (FPR @ 0.05)} & 20.8 & 4.6  & 22   & 7.1 & 7.9  & 12.3 & 0    & 1.5  & 3.5  & 29.2 & 12.9 & 0   & 2.4  & 58.1 & 16.3  \\
                        & \textbf{TPR (FPR @ 0.1)}  & 25.4 & 14.2 & 28.2 & 7.4 & 13.4 & 13.7 & 0    & 2.4  & 9.1  & 32.9 & 14.4 & 0   & 5.1  & 75.8 & 19    \\ \midrule
\textbf{BIM (16/255)}   & \textbf{TPR (FPR @ 0.01)} & 9.7  & 4    & 11.7 & 5.7 & 2.9  & 6.8  & 0    & 0.1  & 0.5  & 51.8 & 5.1  & 0   & 1.4  & 40.3 & 36.9  \\
                        & \textbf{TPR (FPR @ 0.05)} & 11.6 & 4.1  & 22.4 & 6.6 & 6.5  & 9.2  & 0    & 2.5  & 3.1  & 53   & 5.5  & 0   & 2.3  & 48.5 & 37.9  \\
                        & \textbf{TPR (FPR @ 0.1)}  & 15.7 & 14.1 & 30   & 6.3 & 13.8 & 10   & 0    & 3.4  & 10.5 & 54.4 & 6.8  & 0   & 7    & 62.4 & 39    \\ \midrule
\textbf{BIM (32/255)}   & \textbf{TPR (FPR @ 0.01)} & 8.2  & 4.9  & 12.2 & 2.9 & 1.1  & 4.7  & 0    & 0.2  & 0.5  & 72.3 & 2.2  & 0   & 2    & 21.6 & 57.3  \\
                        & \textbf{TPR (FPR @ 0.05)} & 9.6  & 5    & 25.1 & 3.2 & 12.5 & 6.5  & 0    & 2.5  & 3.5  & 73.2 & 2.9  & 0   & 3.4  & 27.5 & 57.4  \\
                        & \textbf{TPR (FPR @ 0.1)}  & 10.1 & 16   & 26.5 & 3.5 & 26   & 6.8  & 0    & 4.3  & 9    & 74   & 3.9  & 0   & 7.4  & 39.7 & 57.7  \\ \midrule
\textbf{BIM (64/255)}   & \textbf{TPR (FPR @ 0.01)} & 5.1  & 5.7  & 1.5  & 1.4 & 11.6 & 3    & 0    & 0.4  & 0.5  & 81.2 & 1.1  & 0   & 1.1  & 5.4  & 58.8  \\
                        & \textbf{TPR (FPR @ 0.05)} & 5.2  & 6    & 1.2  & 1.8 & 36.7 & 4.3  & 0    & 2.1  & 4.1  & 81.2 & 1.4  & 0   & 3.2  & 7.7  & 59.9  \\
                        & \textbf{TPR (FPR @ 0.1)}  & 5.1  & 39.6 & 8.2  & 2.3 & 55.5 & 6.1  & 0    & 3.6  & 12   & 81.6 & 1.7  & 0   & 8.1  & 13.4 & 59.2  \\ \midrule
\textbf{Auto-PGD (0.15)} & \textbf{TPR (FPR @ 0.01)} & 32.5 & 19.6 & 14.8 & 2.4 & 98.5 & 6.6  & 1.2  & 7.5  & 0.5  & 98   & 2.5  & 1.4 & 5.1  & 1.3  & 97.1  \\
                        & \textbf{TPR (FPR @ 0.05)} & 41.2 & 20.5 & 22.3 & 3.1 & 99.2 & 7    & 1.5  & 12.2 & 4.1  & 98.1 & 2.8  & 1.7 & 17.3 & 1.4  & 97.1  \\
                        & \textbf{TPR (FPR @ 0.1)}  & 51.5 & 100  & 41.5 & 4.2 & 99.7 & 7.4  & 4.9  & 23.2 & 7.9  & 98.2 & 3.3  & 1.8 & 45.9 & 2.8  & 97.1  \\ \midrule
\textbf{CW (0.15)}      & \textbf{TPR (FPR @ 0.01)} & 26.7 & 9.5  & 12.2 & 2.5 & 34.5 & 7.2  & 0    & 0.1  & 0.5  & 45.6 & 10.2 & 0   & 1.1  & 0.8  & 75.2  \\
                        & \textbf{TPR (FPR @ 0.05)} & 28.2 & 9.9  & 20.5 & 3.6 & 67.9 & 10.4 & 0    & 0.4  & 1.1  & 67.3 & 13.5 & 0   & 3.4  & 14.5 & 87.3  \\
                        & \textbf{TPR (FPR @ 0.1)}  & 29.3 & 82.6 & 25.7 & 4.1 & 85.3 & 15.5 & 0    & 1.6  & 8.3  & 76.2 & 16.4 & 0   & 5.4  & 42.3 & 94.1  \\ \bottomrule
\end{tabular}%
}
\end{table*}

\subsection{False positive rates of unsupervised detectors}\label{sec:falsepositiverate}
We evaluate unsupervised detectors with the True Positive Rate (TPR), computed as the ratio of the total number of correctly identified adversarial examples to the overall number of adversarial examples. We compute the TPR of detectors by using the threshold learned during training for specific thresholds on the validation set. However, it is highly unlikely that the detector will get the same FPR on the test set. Hence, computing another metric, false positive rate (FPR), is an important criterion. FPR measures the ratio of the number of natural images identified as adversarial images to the total number of natural images. We compare the FPR of PASA against baselines in Figure \ref{fig:fprplot}, where we plot the average FPRs on the test set across all attacks corresponding to the FPR associated with the threshold learned during training (1\%, 5\%, 10\%). The dotted line represents the ideal position of the plot. Detectors that are closer to this line have lower FPRs and are better detectors in classifying the benign images correctly. We can observe that PASA consistently obtains better false positive rates than other methods on CIFAR-10 and ImageNet. On CIFAR-100, while TWS has the lowest FPRs overall, our method obtains better FPRs than U-LOO and MagNet.

\begin{figure*}
    \centering
    \begin{subfigure}{0.3\textwidth}
        \includegraphics[width=\linewidth]{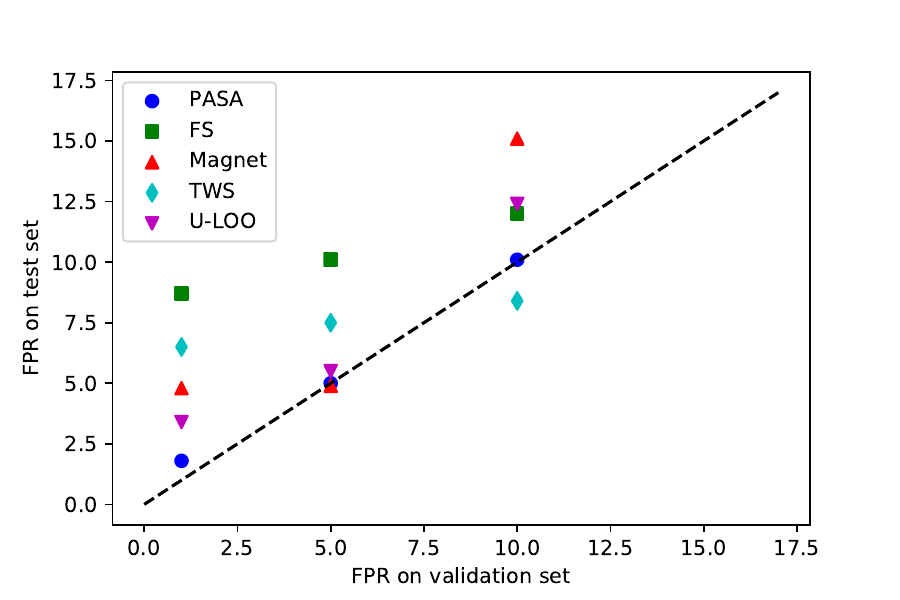}
        \caption{CIFAR-10}
        \label{subfig:a}
    \end{subfigure}
    \begin{subfigure}{0.3\textwidth}
        \includegraphics[width=\linewidth]{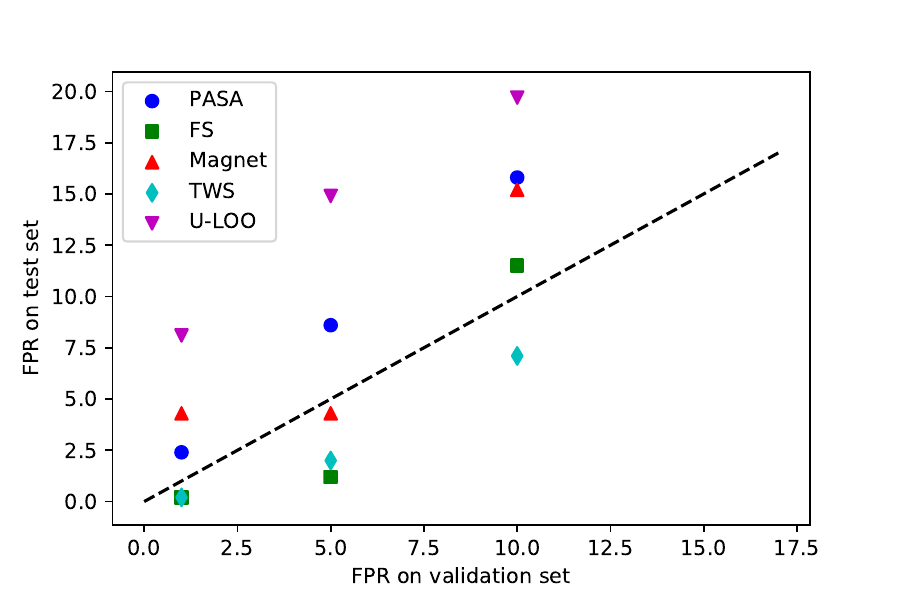}
        \caption{CIFAR-100}
        \label{subfig:b}
    \end{subfigure}
    \begin{subfigure}{0.3\textwidth}
        \includegraphics[width=\linewidth]{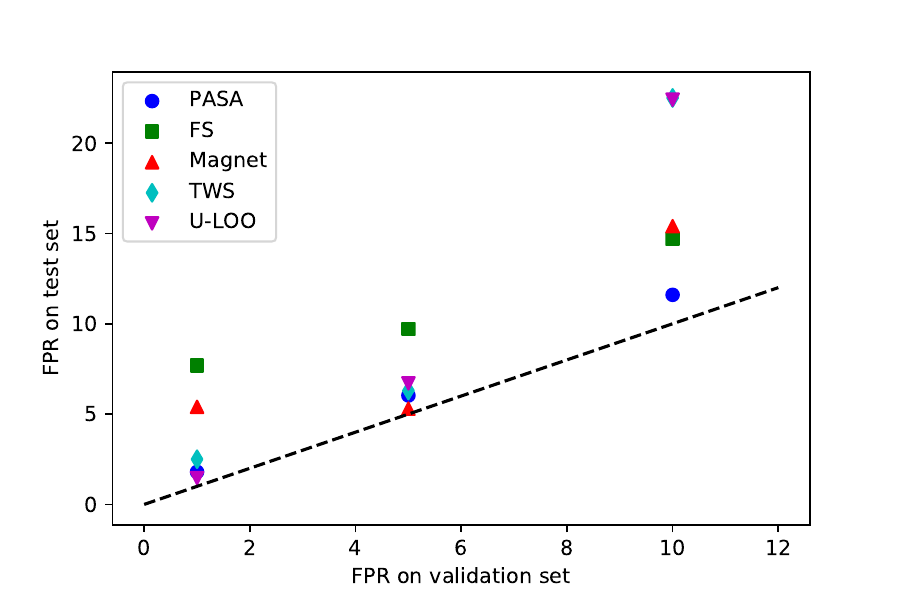}
        \caption{ImageNet}
        \label{subfig:c}
    \end{subfigure}
    \caption{FPR on validation set vs FPR on test set.}
    \label{fig:fprplot}
\end{figure*}

\subsection{Ablation study}\label{sec:ablationstudy}
PASA comprises two statistical metrics: prediction sensitivity and attribution sensitivity. In this ablation study, we assess the individual performance of each metric. Specifically, our focus is on evaluating the detection performance of our proposed method when utilizing only one of the statistical metrics. We maintain a similar experimental setup, selecting 1000 benign images from the test set that are accurately classified by the model. For each attack, we generate 1000 corresponding adversarial images. We use the thresholds for prediction sensitivity and attribution sensitivity learned during the training of our detector.  The collective average AUC for each dataset under different attacks is illustrated in Figure \ref{fig:ablationstudy}. We summarize the results below:

\noindent {1.} On CIFAR-10 (VGG), prediction sensitivity outperforms attribution sensitivity for detection even though both metrics have high performance on average.

\noindent {2.} On CIFAR-10 CIFAR-10 (ResNet), attribution sensitivity outperforms prediction sensitivity. Its standalone performance is almost equivalent to the combined performance.

\noindent {3.} On CIFAR-100, the performance of attribution sensitivity and prediction sensitivity is almost equivalent.

\noindent {4.} On ImageNet, both metrics have lower performance when used standalone. The combined performance is significantly better than the individual metric.

Detailed results can be found in Appendix \ref{appendix:ablation}, where we demonstrate that AS and PS exhibit sensitivity to different attack types, and the combination of both metrics provides a more balanced detection strategy across various attack types.

\subsection{{Evaluation with adaptive attacks}}\label{sec:adaptiveattacks}


In the previous experiments, we assume that the adversary has access to the model details but does not know the details of our detection mechanism. While this is a realistic assumption, it does not provide the robustness measure of the proposed detector. We now evaluate the performance of our proposed method under adaptive attacks to evaluate its robustness. Adaptive attacks are adversarial attacks targeted at a defense mechanism and are adapted to the specific details of a defense. Since our detection approach comprises two statistical measures, we perform adaptive attacks on both components \cite{tramer2020adaptive}. We optimize the PGD attack with perturbation set at $0.1$ and evaluate our results on the CIFAR-10 (ResNet) dataset.

First, we attack the feature attribution method, Integrated Gradient (IG). An adversary tries to deceive both the target classifier and IG. Similar to ADV2 attack \cite{zhang2020interpretable}, this attack generates an adversarial image $\textbf{x}^*$ such that following conditions are satisfied: \textit{1) target classifier ($F$) misclassifies $\textbf{x}^*$}, \textit{2) IG generates attribution similar to benign counterpart $\textbf{x}$ where the similarity is measured using the intersection-over-union (IoU) test, widely used in object detection \cite{he2017p}}, and \textit{3) the difference between benign $\textbf{x}$ and adversarial $\textbf{x}^*$ image is minimized}.

We solve the following optimization for an image $\textbf{x}$,
\begin{equation}\label{eqn:attackIG}
    min_{\textbf{x}^*} L_1(F(\textbf{x}^*), y^*) + c*L_2(IG(\textbf{x}^*), IG(\textbf{x}))
\end{equation}

where $L_1$ is the prediction loss used by PGD attack, 
$L_2=||IG(\textbf{x}^*)-IG(\textbf{x})||_2$ is the loss measuring the difference between the attribution map of the benign image and its adversarial counterpart, and $c$ is a hyper-parameter to balance the two losses. Similar to ADV2 attack \cite{zhang2020interpretable}, we observed that it is inefficient to search for adversarial input by directly running the updates using Eq. \ref{eqn:attackIG}. Hence, we perform a warm start by first running a fixed number of steps for the regular PGD attack and then resume the updates of Eq. \ref{eqn:attackIG}. We use the following values of $c \in [5,10,20,30,50]$ for iteration steps $\in [300, 200, 100, 50]$. We generate 1000 adversarial images according to the attack strategy of Eq. \ref{eqn:attackIG}. These adversarial images obtain an attack success rate of 100\% to fool the model (condition 1); the mean IoU score between benign and adversarial attribution is 43\% (condition 2) (note that ADV2 \cite{zhang2020interpretable} also obtained IoU of only 50\% on attacking Vanilla Gradient \cite{sundararajan2017axiomatic}). The mean L2 distortion of successful adaptive adversarial images is 2.8, which is slightly higher compared with the PGD attack (2.6) (condition 3). We apply our detection strategy on the adversarial images and obtain an AUC score of $0.75$. The adaptive attack takes a significantly longer time compared with PGD. A normal PGD attack takes about 0.065 seconds to generate an adversarial sample for a single image, whereas this adaptive attack takes around 26.21 seconds, which is $\sim$400 times slower.

\begin{table}[]
\centering
\caption{Performance of PASA against Adaptive Attacks. Complexity measures computation time in seconds.}
\label{tab:adaptiveattack}
\resizebox{0.47\textwidth}{!}{%
\begin{tabular}{@{}lcccccc@{}}
\toprule
\multicolumn{1}{c}{\textbf{Attack}} &
  \multicolumn{2}{c}{\textbf{Attack success rate}} &
  \multicolumn{2}{c}{\textbf{Detection AUC}} &
  \multicolumn{2}{c}{\textbf{Complexity (time)}} \\ \midrule
\textbf{} &
  \multicolumn{1}{c}{\textbf{Before}} &
  \multicolumn{1}{c}{\textbf{After}} &
  \multicolumn{1}{c}{\textbf{Before}} &
  \multicolumn{1}{c}{\textbf{After}} &
  \multicolumn{1}{c}{\textbf{Before}} &
  \multicolumn{1}{c}{\textbf{After}} \\ \midrule
\textbf{Attack on IG}     & 100\% & 100\% & 0.98 & 0.75   & 0.065 & 26.21 \\
\textbf{Attack on logits} & 100\% & 100\% & 0.98 & 0.76   & 0.065 & 0.32 \\
\textbf{Combined attack}  & 100\% & 100\% & 0.98 & 0.69   & 0.065 & 28.33 \\ \bottomrule
\end{tabular}%
}
\end{table}

Next, we perform an adaptive attack on the model logits. The adversary creates adversarial images in such a way that the distribution of logits is closer to the logits of benign images. We follow the logit matching attack of Tramer et al. \cite{tramer2020adaptive}. We solve the following optimization to obtain an adversarial image $\textbf{x}^*$
\begin{equation}\label{eqn:attacklogit}
    min_{\textbf{x}^*} L_1(F(\textbf{x}^*), y^*) + L_3(Z(\textbf{x}^*), Z(\textbf{x}))
\end{equation}

where $L_1$ is prediction loss of a PGD attack. The second loss term $L_3=||Z(\textbf{x}^*)-Z(\textbf{x})||_2$ is the mean square loss between logits of adversarial and benign images. The attack runs for a fixed number of iterations (given by PGD iterations) and produces adversarial samples whose logits are closer to benign counterparts. For 1000 adversarial images obtained using this strategy, the adaptive attack still achieves a 100\% attack success rate. The mean L2 distortion of successful samples is 2.7, which is similar to the PGD attack (2.6). We obtain an AUC score of  $0.76$ for the detector. We observe that attacking only one component of our detector does not significantly impact the overall detection.

\begin{figure}
    \centering
    \includegraphics[width=.25\textwidth]{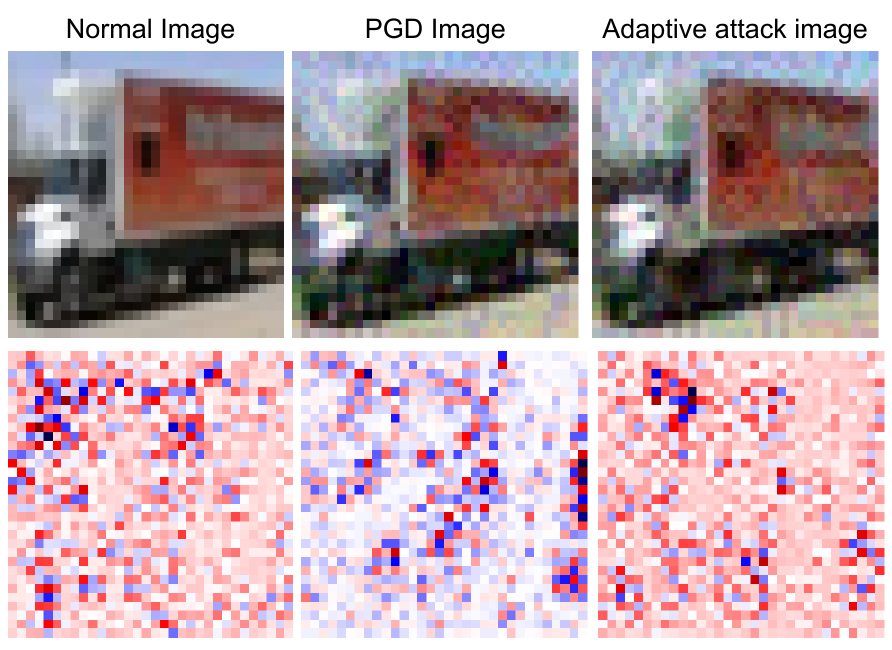}
    \caption{Attribution map (second row) for three different images (first row): benign image, adversarial image from PGD attack, and adversarial image from adaptive attack.}
    \label{fig:adaptiveattack}
\end{figure}

Finally, we introduce attacks on both the attribution method and model logits. We introduce two different losses and solve the following optimization to obtain an adversarial image $\textbf{x}^*$ for an input image $\textbf{x}$
\begin{equation}\label{eqn:attackIGandLogit}
    min_{\textbf{x}^*} L_1(.) + c*L_2(.) + L_3(.)
\end{equation}

where $L_1(F(\textbf{x}^*), y*)$ is prediction loss, $L_2(IG(\textbf{x}^*), IG(\textbf{x}))$ is the loss measuring the difference between the attribution map of benign samples and their adversarial counterparts, $L_3(Z(\textbf{x}^*), Z(\textbf{x}))$ measures the loss between logits of benign and adversarial samples and $c$ is a hyper-parameter. We perform a warm start by searching for adversarial samples using PGD attack and then iteratively optimize Eq. \ref{eqn:attackIGandLogit} to obtain adversarial samples with attribution vector and logits similar to benign samples. For a similar test setting, the adaptive attack obtains an attack success rate of 100\%. The mean L2 distortion of successful samples is 2.7. The AUC score of the detector is now reduced to $0.69$. This adaptive attack takes around 28.33 seconds on average for each sample. {Figure \ref{fig:adaptiveattack} shows the results of this adaptive attack. The first row shows images from class ``Truck" from CIFAR-10, and the second row shows its heatmap computed using IG. We can observe that the attribution map of a PGD image differs significantly from the attribution map of a normal image. After performing an adaptive attack (which attacks both feature attribution and model logit), the adversary obtains a perturbed image with its attribution map similar to that of a natural image.}

We summarize the result in Table \ref{tab:adaptiveattack} where attack success rate measures the success of the attack in changing the label of an image, detection AUC measures the performance of our detector in detecting adversarial images, and complexity measures the time required by the attack for a single image (in seconds). We observe that performing adaptive attacks against both components of our detector increases computational complexity. However, even though the detector performance drops with adaptive attacks, PASA is still able to achieve a competitive performance under this strongest adversary assumption. In Table \ref{tab:my-adaptive-attacks}, we evaluate the performance of different detection methods against the adaptive adversarial samples obtained using Eqn. \ref{eqn:attackIGandLogit}. This adversarial attack was specifically designed to evade PASA's detection mechanism. However, all detection methods have considerably lower AUCs in detecting the adversarial samples.

\subsubsection{Analysis} Prior works have shown that it is possible to add imperceptible perturbation to images for generating random attribution maps \cite{ghorbani2019interpretation, zhang2020interpretable, subramanya2018towards}. However, evading a classifier and generating an attribution similar to benign counterparts is much more challenging. Attacks like ADV2 \cite{zhang2020interpretable} only achieved a 50\% IOU when targeting the Vanilla Gradient method \cite{simonyan2013deep}. In our evaluation, the adaptive attack only achieved 43\% IOU on attacking the Integrated Gradient (IG) method. This means there is still a significant discrepancy between the attribution map of benign and adversarial samples that PASA can utilize in detection. This difficulty stems from the challenge of satisfying two counter-intuitive objectives: retaining the adversarial label while aligning the attribution with the benign example. This was validated by a recent work \cite{boopathy2020proper}, which shows that it is difficult to remove the $L_1$-norm of attribution discrepancy between benign and adversarial images when an attribution method satisfying completeness axiom (e.g. IG \cite{sundararajan2017axiomatic}) is used. These findings suggest that an explanation method like IG can help detect discrepancies between benign and adversarial examples.

\begin{table}[]
\centering
\caption{Evaluation of Adaptive Attacks}
\label{tab:my-adaptive-attacks}
\resizebox{0.38\textwidth}{!}{%
\begin{tabular}{@{}lccccc@{}}
\toprule
\textbf{Method}       & \textbf{FS} & \textbf{MagNet} & \textbf{TWS} & \textbf{U-LOO} & \textbf{PASA} \\ \midrule
\textbf{AUC (Before)} &    0.13  &    0.91 &    0.14 &   0.68    &   0.98    \\
\textbf{AUC (After)}  & 0.54    & 0.51     & 0.35   & 0.59  &     0.69          \\ \bottomrule
\end{tabular}%
}
\end{table}

\section{Application On Security Dataset}\label{sec:securitydataset}

In this section, we demonstrate the application of PASA on a network security dataset. We specifically focus on a network intrusion detection problem and evaluate PASA using the updated CIC-IDS2017 dataset \cite{engelen2021troubleshooting}. Since the goal of an adversarial attack on network data is to classify attack samples as benign, we preprocess the data by assigning a single attack label to different attack-traffic types. Subsequently, we build a multi-layer perceptron model with a binary classification objective, which achieves an accuracy of 99.04\% on the test set. Further details about the dataset, and model can be found in Appendix \ref{appendix:dataset} and \ref{apendix:networkintrusion}. 

We generate adversarial samples using FGSM, BIM, PGD, CW and Auto-PGD attacks by considering a threat model where an adversary is able to manipulate the dataset in feature-space. Hence, such attacks might not be representative of a realistic settings and might produce ``unrealizable" adversarial samples \cite{sheatsley2021robustness, apruzzese2022modeling}. 

In Table \ref{tab:ndss}, we compare U-LOO and PASA (AUC scores) on the intrusion detection model. It is important to note that all baseline methods we considered cannot be directly applied to security applications. For instance, FS \cite{xu2017feature} is designed for image data and necessitates the application of various image filters for adversarial detection. MagNet  \cite{meng2017magnet}, proposed for image datasets, relies on an auto-encoder model to compute the reconstruction error. TWS  \cite{hu2019new} leverages the difference in softmax predictions between the original and noisy inputs to detect adversarial samples. However, there is a negligible difference between the softmax outputs of the original and noisy inputs in the binary classification task, rendering TWS ineffective. On the other hand, U-LOO works well in our scenario as it measures the interquartile range (IQR) of feature attributions for both benign and adversarial samples.

\section{{Discussion}}\label{sec:discussion}
\subsection{Indicators of attack failures}
Indicators of attack failures \cite{pintor2022indicators} serve to uncover potential vulnerabilities in the adversarial robustness evaluation.  In our defense assessment, two specific scenarios merit consideration: \textit{a) Non-converging attack:} Prior defenses \cite{buckman2018thermometer, pang2019improving} often employed small steps in PGD attack, introducing the risk of non-convergence. As discussed in Sec. \ref{sec:attackdetail}, we mitigated this risk by opting for larger iteration steps (40) in our PGD attack formulation. Aligning with Pintor et al.\cite{pintor2022indicators} recommendation, we reassessed detection methods on the PGD attack with 100 steps. Notably, all detection methods exhibited no significant performance degradation when compared with the previous evaluation of PGD samples (steps=40). Results are summarized in Appendix \ref{appendix:pgd100steps}. \textit{b. Non-adaptive attacks:} Prior defenses \cite{das2018shield} have developed adaptive attacks, neglecting non-differentiable or additional defense components. However, adaptive attacks that ignore a component of a defense do not guarantee successful bypassing of the target model. In our case, both model prediction and feature attribution are differentiable and we incorporate them in crafting adaptive adversarial samples, as discussed in Section \ref{sec:adaptiveattacks}.

\begin{table}[]
\centering
\caption{Evaluation on updated CIC-IDS2017}
\label{tab:ndss}
\resizebox{0.46\textwidth}{!}{%
\begin{tabular}{@{}lccccc@{}}
\toprule
\multicolumn{1}{c}{\textbf{Method}} & \textbf{FGSM} & \textbf{PGD} & \textbf{BIM} & \textbf{CW} & \textbf{Auto-PGD} \\ \midrule
\textbf{PASA} & 0.99 $\pm$ 0.00     & 0.98 $\pm$ 0.02    & 0.99 $\pm$ 0.01    & 0.98 $\pm$ 0.03 & 0.97 $\pm$ 0.03  \\
\textbf{U-LOO \cite{yang2020ml}}    & 0.70 $\pm$ 0.05     & 0.66 $\pm$ 0.07    & 0.76 $\pm$ 0.01    & 0.75 $\pm$ 0.01  & 0.63 $\pm$ 0.00   \\ \bottomrule
\end{tabular}%
}
\end{table}

\subsection{Efficient Lightweight detection} 
The main strength of PASA is its lightweight detection approach. Since it requires the computation of two statistics by probing a given input image with noise, it has low inference latency. The simplicity of this approach means that it has a small memory footprint, making it suitable for deployment on resource-constrained devices or in scenarios where computational resources are limited. While other unsupervised methods like MagNet \cite{meng2017magnet}, FS \cite{xu2017feature}, and TWS \cite{hu2019new}, also have low inference time, PASA outperforms these methods in detecting attacks against CIFAR and ImageNet models. We evaluated the inference time, training time, and memory usage (peak memory required) of different detection methods for 1000 images and report average results in Table \ref{tab:my-memory}, where we can observe that PASA is faster than LOO but slower than TWS, FS, and MagNet. MagNet requires the highest training time, attributed to the necessity of training a separate defense model. PASA has a moderate training time, significantly lower than MagNet and LOO. PASA also has a moderate memory usage, higher than TWS but lower than LOO, and MagNet.

\subsection{On explanation methods and their explanation fragility}

An adversary can introduce perturbations to an input sample such that it is misclassified, but its attribution map is similar to the attribution of benign sample \cite{zhang2020interpretable}. This is an attack on the explanation method. Such attacks affect the performance of detection methods that utilize disparity in feature attribution. However, the success of this attack depends on the attribution method. In our approach, we employed IG\cite{sundararajan2017axiomatic}, which displayed higher resilience against adaptive adversarial attacks. A similar result was demonstrated in a study by Vardhan et al. \cite{vardhan2021exad}. However, the Vanilla Gradient \cite{simonyan2013deep} is more sensitive to such attacks, producing similar attribution maps between benign and adversarial samples \cite{zhang2020interpretable}. Alternative feature attribution methods such as LRP \cite{bach2015pixel} and GBP \cite{springenberg2014striving} can also be manipulated to produce targeted attribution maps mimicking those of benign samples, effectively fooling the detector \cite{vardhan2021exad}.

\begin{table}[]
\centering
\caption{Computational overhead of detection methods}
\label{tab:my-memory}
\resizebox{0.44\textwidth}{!}{%
\begin{tabular}{@{}lccc@{}}
\toprule
\multicolumn{1}{c}{\textbf{Method}} & \textbf{Inference Time (s)} & \textbf{Training Time (s)} & \textbf{Memory Usage (MB)} \\ \midrule
\textbf{PASA}   & 0.0156 & 15.460   & 2701.46 \\
\textbf{U-LOO  \cite{yang2020ml}}    & 0.0540 & 53.076   & 2800.62 \\
\textbf{FS \cite{xu2017feature}}     & 0.0085 & 8.456    & 2644.98 \\
\textbf{MagNet \cite{meng2017magnet}} & 0.0007 & 1262.778 & 2744.41 \\
\textbf{TWS  \cite{hu2019new}}    & 0.0051 & 5.406    & 2576.08 \\ \bottomrule
\end{tabular}%
}
\end{table}


\subsection{Utilizing latent representations} 
PASA utilizes the final layer features from the classification model (logit layer) and the explanation method (input attribution) for detecting adversarial samples. This can be extended to incorporate features from multiple intermediate layers as well. Few recent works exploit the behavior of neural networks to benign and adversarial samples in hidden layers to design supervised adversarial detectors \cite{yang2020ml, sperl2020dla}. However, since DNNs can have many hidden layers, it will require careful consideration to include features from specific layers; otherwise, the detectors may overfit the training data due to a large number of features. We will explore the inclusion of statistics from latent representations in our future work.

\subsection{Limitations}

PASA works well for detecting adversarial samples generated through $L_\infty$ attacks. Such attacks aim to maximize perturbation within a bounded norm. PASA is effective against them because it can capture significant changes in attribution and prediction differences, which result from substantial perturbations. In contrast, $L_1, L_2$ attacks make minimal changes to the input by altering only a few pixels and minimizing perturbation magnitude, respectively (See Table \ref{tab:distortion}). Integrated Gradient may not capture such small or subtle perturbations effectively. For instance, evaluating the detection methods on $L_2$ PGD attacks on the CIFAR-10 ResNet at $\epsilon=64/255$, we obtained the following AUC: 0.59 (PASA), 0.55 (FS), MagNet (0.51), U-LOO (0.52), TWS (0.38). 

Other attacks beyond the $L_p$ attack (e.g., patch attacks) modify only a specific part of an image by adding a patch, so directly applying PASA and observing the difference in prediction and attribution does not work. However, such attacks still perform significant modifications to hidden feature maps that produce changes in model prediction. Our future work will focus on utilizing noise-based approaches on hidden-activations in detecting other evasion attacks of $L_1$, $L_2$ norms, and patch attacks.

PASA also has a noticeable drop in performance with images of lower resolution like MNIST. This is because the granularity at which IG attributes importance to individual features in MNIST is smaller and hence, it has lower sensitivity to noise, the quality utilized by PASA in detection. 

Like other noise-based approaches \cite{hu2019new, roth2019odds}, our noise parameter needs to be determined empirically. This means that the effectiveness of the method can depend on the specific dataset and problem at hand. Selecting the optimal noise parameter requires experimentation, which could be time-consuming before deployment. However, we have demonstrated through different datasets and network architectures that once the optimal noise value is discovered, PASA can be generalized across datasets for lightweight detection of attacks.

\begin{table}[]
\centering
\caption{Average distortion between benign \& adversarial CIFAR-10 images at different attack strength}
\label{tab:distortion}
\resizebox{0.24\textwidth}{!}{%
\begin{tabular}{@{}lcccc@{}}
\toprule
\textbf{Attack} &
  \multicolumn{1}{l}{\textbf{8/255}} &
  \multicolumn{1}{l}{\textbf{16/255}} &
  \multicolumn{1}{l}{\textbf{32/255}} &
  \multicolumn{1}{l}{\textbf{64/255}} \\ \midrule
$L_1$ PGD &
  0.001 &
  0.002 &
  0.003 &
  0.007 \\
  $L_2$ PGD &
  0.031 &
  0.063 &
  0.125 &
  0.251 \\
$L_\infty$ PGD &
  1.456 &
  2.706 &
  4.818 &
  8.512 \\ \bottomrule
\end{tabular}%
}
\end{table}

\section{Conclusion}\label{sec:conclusion}
In this paper, we propose PASA, a lightweight attack-agnostic, unsupervised method for detecting adversarial samples. We use noise as a probing tool to measure the sensitivity of model prediction and feature attribution. We learn thresholds of sensitivity scores from benign samples and utilize them for detection. PASA outperforms existing statistical unsupervised detectors in classifying adversarial samples on the updated CIC-IDS2017, CIFAR-10, CIFAR-100, and ImageNet datasets. PASA displays robust performance in detecting adversarial samples even when an adversary has full knowledge of the detector. We aim to extend the scope of our approach in future studies, particularly to detect adversarial attacks of the $L_0$ and $L_2$ norm and physically realizable patch-based attacks, and improve the security of diverse systems that use deep learning models.

\section*{Acknowledgement}
\noindent 
We are grateful to anonymous reviewers at IEEE Euro S\&P for their valuable feedback. This work was supported by Toyota InfoTech Labs through Unrestricted Research Funds and RIT AI Seed Fund Program.

\bibliographystyle{plain}
\bibliography{main}

\begin{thebibliography}{10}

\bibitem{aigrain2019detecting}
Jonathan Aigrain and Marcin Detyniecki.
\newblock Detecting adversarial examples and other misclassifications in neural networks by introspection.
\newblock {\em CoRR}, abs/1905.09186, 2019.

\bibitem{anderson2018ember}
Hyrum~S Anderson and Phil Roth.
\newblock Ember: an open dataset for training static pe malware machine learning models.
\newblock {\em arXiv preprint arXiv:1804.04637}, 2018.

\bibitem{apruzzese2022modeling}
Giovanni Apruzzese, Mauro Andreolini, Luca Ferretti, Mirco Marchetti, and Michele Colajanni.
\newblock Modeling realistic adversarial attacks against network intrusion detection systems.
\newblock {\em Digital Threats: Research and Practice (DTRAP)}, 3(3):1--19, 2022.

\bibitem{bach2015pixel}
Sebastian Bach, Alexander Binder, Gr{\'e}goire Montavon, Frederick Klauschen, Klaus-Robert M{\"u}ller, and Wojciech Samek.
\newblock On pixel-wise explanations for non-linear classifier decisions by layer-wise relevance propagation.
\newblock {\em PloS one}, 10(7):e0130140, 2015.

\bibitem{Bhusal2022SoKME}
Dipkamal Bhusal, Rosalyn Shin, Ajay~Ashok Shewale, Monish Kumar~Manikya Veerabhadran, Michael Clifford, Sara Rampazzi, and Nidhi Rastogi.
\newblock Sok: Modeling explainability in security analytics for interpretability, trustworthiness, and usability.
\newblock In {\em The 18th International Conference on Availability, Reliability and Security (ARES 2023)}, 2023.

\bibitem{biggio2018wild}
Battista Biggio and Fabio Roli.
\newblock Wild patterns: Ten years after the rise of adversarial machine learning.
\newblock In {\em Proceedings of the 2018 ACM SIGSAC Conference on Computer and Communications Security}, pages 2154--2156, 2018.

\bibitem{bojarski2016end}
Mariusz Bojarski, Davide~Del Testa, Daniel Dworakowski, Bernhard Firner, Beat Flepp, Prasoon Goyal, Lawrence~D. Jackel, Mathew Monfort, Urs Muller, Jiakai Zhang, Xin Zhang, Jake Zhao, and Karol Zieba.
\newblock End to end learning for self-driving cars.
\newblock {\em CoRR}, abs/1604.07316, 2016.

\bibitem{boopathy2020proper}
Akhilan Boopathy, Sijia Liu, Gaoyuan Zhang, Cynthia Liu, Pin-Yu Chen, Shiyu Chang, and Luca Daniel.
\newblock Proper network interpretability helps adversarial robustness in classification.
\newblock In {\em International Conference on Machine Learning}, pages 1014--1023. PMLR, 2020.

\bibitem{buckman2018thermometer}
Jacob Buckman, Aurko Roy, Colin Raffel, and Ian Goodfellow.
\newblock Thermometer encoding: One hot way to resist adversarial examples.
\newblock In {\em International conference on learning representations}, 2018.

\bibitem{bykov2022noisegrad}
Kirill Bykov, Anna Hedstr{\"o}m, Shinichi Nakajima, and Marina M-C H{\"o}hne.
\newblock Noisegrad—enhancing explanations by introducing stochasticity to model weights.
\newblock In {\em Proceedings of the AAAI Conference on Artificial Intelligence}, volume~36, pages 6132--6140, 2022.

\bibitem{carlini2017adversarial}
Nicholas Carlini and David Wagner.
\newblock Adversarial examples are not easily detected: Bypassing ten detection methods.
\newblock In {\em Proceedings of the 10th ACM workshop on artificial intelligence and security}, pages 3--14, 2017.

\bibitem{carlini2017towards}
Nicholas Carlini and David Wagner.
\newblock Towards evaluating the robustness of neural networks.
\newblock In {\em 2017 ieee symposium on security and privacy (sp)}, pages 39--57. Ieee, 2017.

\bibitem{chalasani2020concise}
Prasad Chalasani, Jiefeng Chen, Amrita~Roy Chowdhury, Xi~Wu, and Somesh Jha.
\newblock Concise explanations of neural networks using adversarial training.
\newblock In {\em International Conference on Machine Learning}, pages 1383--1391. PMLR, 2020.

\bibitem{croce2020reliable}
Francesco Croce and Matthias Hein.
\newblock Reliable evaluation of adversarial robustness with an ensemble of diverse parameter-free attacks.
\newblock In {\em International conference on machine learning}, pages 2206--2216. PMLR, 2020.

\bibitem{das2018shield}
Nilaksh Das, Madhuri Shanbhogue, Shang-Tse Chen, Fred Hohman, Siwei Li, Li~Chen, Michael~E Kounavis, and Duen~Horng Chau.
\newblock Shield: Fast, practical defense and vaccination for deep learning using jpeg compression.
\newblock In {\em Proceedings of the 24th ACM SIGKDD International Conference on Knowledge Discovery \& Data Mining}, pages 196--204, 2018.

\bibitem{davis2006relationship}
Jesse Davis and Mark Goadrich.
\newblock The relationship between precision-recall and roc curves.
\newblock In {\em Proceedings of the 23rd international conference on Machine learning}, pages 233--240, 2006.

\bibitem{deng2009imagenet}
Jia Deng, Wei Dong, Richard Socher, Li-Jia Li, Kai Li, and Li~Fei-Fei.
\newblock Imagenet: A large-scale hierarchical image database.
\newblock In {\em 2009 IEEE conference on computer vision and pattern recognition}, pages 248--255. Ieee, 2009.

\bibitem{engelen2021troubleshooting}
Gints Engelen, Vera Rimmer, and Wouter Joosen.
\newblock Troubleshooting an intrusion detection dataset: the cicids2017 case study.
\newblock In {\em 2021 IEEE Security and Privacy Workshops (SPW)}, pages 7--12. IEEE, 2021.

\bibitem{feinman2017detecting}
Reuben Feinman, Ryan~R. Curtin, Saurabh Shintre, and Andrew~B. Gardner.
\newblock Detecting adversarial samples from artifacts.
\newblock {\em CoRR}, abs/1703.00410, 2017.

\bibitem{gardner2015deep}
Jacob~R. Gardner, Matt~J. Kusner, Yixuan Li, Paul Upchurch, Kilian~Q. Weinberger, and John~E. Hopcroft.
\newblock Deep manifold traversal: Changing labels with convolutional features.
\newblock {\em CoRR}, abs/1511.06421, 2015.

\bibitem{ghorbani2019interpretation}
Amirata Ghorbani, Abubakar Abid, and James Zou.
\newblock Interpretation of neural networks is fragile.
\newblock In {\em Proceedings of the AAAI conference on artificial intelligence}, volume~33, pages 3681--3688, 2019.

\bibitem{goodfellow2015explaining}
Ian~J. Goodfellow, Jonathon Shlens, and Christian Szegedy.
\newblock Explaining and harnessing adversarial examples.
\newblock In Yoshua Bengio and Yann LeCun, editors, {\em 3rd International Conference on Learning Representations, {ICLR} 2015, San Diego, CA, USA, May 7-9, 2015, Conference Track Proceedings}, 2015.

\bibitem{he2017p}
Kaiming He and Georgia Gkioxari.
\newblock P. doll ar, and r. girshick,“mask r-cnn,”.
\newblock In {\em Proc. IEEE Int. Conf. Comput. Vis}, pages 2980--2988, 2017.

\bibitem{he2015delving}
Kaiming He, Xiangyu Zhang, Shaoqing Ren, and Jian Sun.
\newblock Delving deep into rectifiers: Surpassing human-level performance on imagenet classification.
\newblock In {\em Proceedings of the IEEE international conference on computer vision}, pages 1026--1034, 2015.

\bibitem{he2016deep}
Kaiming He, Xiangyu Zhang, Shaoqing Ren, and Jian Sun.
\newblock Deep residual learning for image recognition.
\newblock In {\em Proceedings of the IEEE conference on computer vision and pattern recognition}, pages 770--778, 2016.

\bibitem{hendrycks2016baseline}
Dan Hendrycks and Kevin Gimpel.
\newblock A baseline for detecting misclassified and out-of-distribution examples in neural networks.
\newblock {\em International Conference on Learning Representations (ICLR)}, 2016.

\bibitem{hosseini2019odds}
Hossein Hosseini, Sreeram Kannan, and Radha Poovendran.
\newblock Are odds really odd? bypassing statistical detection of adversarial examples.
\newblock {\em arXiv preprint arXiv:1907.12138}, 2019.

\bibitem{hu2019new}
Shengyuan Hu, Tao Yu, Chuan Guo, Wei{-}Lun Chao, and Kilian~Q. Weinberger.
\newblock A new defense against adversarial images: Turning a weakness into a strength.
\newblock In {\em Advances in Neural Information Processing Systems 32: Annual Conference on Neural Information Processing Systems 2019, NeurIPS 2019, December 8-14, 2019, Vancouver, BC, Canada}, pages 1633--1644, 2019.

\bibitem{kaggle}
Kaggle.
\newblock Imagenet 1000 (mini).
\newblock https://rb.gy/udu0a, 2020.

\bibitem{kokhlikyan2020captum}
Narine Kokhlikyan, Vivek Miglani, Miguel Martin, Edward Wang, Bilal Alsallakh, Jonathan Reynolds, Alexander Melnikov, Natalia Kliushkina, Carlos Araya, Siqi Yan, and Orion Reblitz-Richardson.
\newblock Captum: A unified and generic model interpretability library for pytorch, 2020.

\bibitem{krizhevsky2009learning}
Alex Krizhevsky, Geoffrey Hinton, et~al.
\newblock Learning multiple layers of features from tiny images.
\newblock 2009.

\bibitem{kumar2020black}
K~Naveen Kumar, C~Vishnu, Reshmi Mitra, and C~Krishna Mohan.
\newblock Black-box adversarial attacks in autonomous vehicle technology.
\newblock In {\em 2020 IEEE Applied Imagery Pattern Recognition Workshop (AIPR)}, pages 1--7. IEEE, 2020.

\bibitem{kurakin2018adversarial}
Alexey Kurakin, Ian~J Goodfellow, and Samy Bengio.
\newblock Adversarial examples in the physical world.
\newblock In {\em Artificial intelligence safety and security}, pages 99--112. Chapman and Hall/CRC, 2018.

\bibitem{lecun1998gradient}
Yann LeCun, L{\'e}on Bottou, Yoshua Bengio, and Patrick Haffner.
\newblock Gradient-based learning applied to document recognition.
\newblock {\em Proceedings of the IEEE}, 86(11):2278--2324, 1998.

\bibitem{li2016understanding}
Jiwei Li, Will Monroe, and Dan Jurafsky.
\newblock Understanding neural networks through representation erasure.
\newblock {\em CoRR}, abs/1612.08220, 2016.

\bibitem{liu2016delving}
Yanpei Liu, Xinyun Chen, Chang Liu, and Dawn Song.
\newblock Delving into transferable adversarial examples and black-box attacks.
\newblock In {\em International Conference on Learning Representations}, 2016.

\bibitem{lundberg2017unified}
Scott~M. Lundberg and Su{-}In Lee.
\newblock A unified approach to interpreting model predictions.
\newblock In Isabelle Guyon, Ulrike von Luxburg, Samy Bengio, Hanna~M. Wallach, Rob Fergus, S.~V.~N. Vishwanathan, and Roman Garnett, editors, {\em Advances in Neural Information Processing Systems 30: Annual Conference on Neural Information Processing Systems 2017, December 4-9, 2017, Long Beach, CA, {USA}}, pages 4765--4774, 2017.

\bibitem{ma2019nic}
Shiqing Ma, Yingqi Liu, Guanhong Tao, Wen-Chuan Lee, and Xiangyu Zhang.
\newblock Nic: Detecting adversarial samples with neural network invariant checking.
\newblock In {\em 26th Annual Network And Distributed System Security Symposium (NDSS 2019)}. Internet Soc, 2019.

\bibitem{ma2018characterizing}
Xingjun Ma, Bo~Li, Yisen Wang, Sarah~M Erfani, Sudanthi Wijewickrema, Grant Schoenebeck, Dawn Song, Michael~E Houle, and James Bailey.
\newblock Characterizing adversarial subspaces using local intrinsic dimensionality.
\newblock In {\em International Conference on Learning Representations}, 2018.

\bibitem{madry2017towards}
Aleksander Madry, Aleksandar Makelov, Ludwig Schmidt, Dimitris Tsipras, and Adrian Vladu.
\newblock Towards deep learning models resistant to adversarial attacks.
\newblock In {\em 6th International Conference on Learning Representations, {ICLR} 2018, Vancouver, BC, Canada, April 30 - May 3, 2018, Conference Track Proceedings}. OpenReview.net, 2018.

\bibitem{meng2017magnet}
Dongyu Meng and Hao Chen.
\newblock Magnet: a two-pronged defense against adversarial examples.
\newblock In {\em Proceedings of the 2017 ACM SIGSAC conference on computer and communications security}, pages 135--147, 2017.

\bibitem{art2018}
Maria-Irina Nicolae, Mathieu Sinn, Minh~Ngoc Tran, Beat Buesser, Ambrish Rawat, Martin Wistuba, Valentina Zantedeschi, Nathalie Baracaldo, Bryant Chen, Heiko Ludwig, Ian Molloy, and Ben Edwards.
\newblock Adversarial robustness toolbox v1.2.0.
\newblock {\em CoRR}, 1807.01069, 2018.

\bibitem{pang2019improving}
Tianyu Pang, Kun Xu, Chao Du, Ning Chen, and Jun Zhu.
\newblock Improving adversarial robustness via promoting ensemble diversity.
\newblock In {\em International Conference on Machine Learning}, pages 4970--4979. PMLR, 2019.

\bibitem{pintor2022indicators}
Maura Pintor, Luca Demetrio, Angelo Sotgiu, Ambra Demontis, Nicholas Carlini, Battista Biggio, and Fabio Roli.
\newblock Indicators of attack failure: Debugging and improving optimization of adversarial examples.
\newblock {\em Advances in Neural Information Processing Systems}, 35:23063--23076, 2022.

\bibitem{ribeiro2016should}
Marco~Tulio Ribeiro, Sameer Singh, and Carlos Guestrin.
\newblock " why should i trust you?" explaining the predictions of any classifier.
\newblock In {\em 22nd ACM SIGKDD}, 2016.

\bibitem{roth2019odds}
Kevin Roth, Yannic Kilcher, and Thomas Hofmann.
\newblock The odds are odd: A statistical test for detecting adversarial examples.
\newblock In {\em International Conference on Machine Learning}, pages 5498--5507. PMLR, 2019.

\bibitem{rumelhart1985llearning}
David~E Rumelhart, Geoffrey~E Hinton, and Ronald~J Williams.
\newblock {\l}learning internal representations by error propagation, {\v{z}} california univ san diego la jolla inst for cognitive science.
\newblock Technical report, Tech. Rep, 1985.

\bibitem{sandler2018mobilenetv2}
Mark Sandler, Andrew Howard, Menglong Zhu, Andrey Zhmoginov, and Liang-Chieh Chen.
\newblock Mobilenetv2: Inverted residuals and linear bottlenecks.
\newblock In {\em Proceedings of the IEEE conference on computer vision and pattern recognition}, pages 4510--4520, 2018.

\bibitem{shafahiadversarial}
Ali Shafahi, W.~Ronny Huang, Christoph Studer, Soheil Feizi, and Tom Goldstein.
\newblock Are adversarial examples inevitable?
\newblock In {\em 7th International Conference on Learning Representations, {ICLR} 2019, New Orleans, LA, USA, May 6-9, 2019}. OpenReview.net, 2019.

\bibitem{sharafaldin2018toward}
Iman Sharafaldin, Arash~Habibi Lashkari, Ali~A Ghorbani, et~al.
\newblock Toward generating a new intrusion detection dataset and intrusion traffic characterization.
\newblock {\em ICISSp}, 1:108--116, 2018.

\bibitem{sheatsley2021robustness}
Ryan Sheatsley, Blaine Hoak, Eric Pauley, Yohan Beugin, Michael~J Weisman, and Patrick McDaniel.
\newblock On the robustness of domain constraints.
\newblock In {\em Proceedings of the 2021 ACM SIGSAC conference on computer and communications security}, pages 495--515, 2021.

\bibitem{shen2017deep}
Dinggang Shen, Guorong Wu, and Heung-Il Suk.
\newblock Deep learning in medical image analysis.
\newblock {\em Annual review of biomedical engineering}, 19:221--248, 2017.

\bibitem{simonyan2013deep}
Karen Simonyan, Andrea Vedaldi, and Andrew Zisserman.
\newblock Deep inside convolutional networks: Visualising image classification models and saliency maps.
\newblock In Yoshua Bengio and Yann LeCun, editors, {\em 2nd International Conference on Learning Representations, {ICLR} 2014, Banff, AB, Canada, April 14-16, 2014, Workshop Track Proceedings}, 2014.

\bibitem{simonyan2014very}
Karen Simonyan and Andrew Zisserman.
\newblock Very deep convolutional networks for large-scale image recognition.
\newblock In Yoshua Bengio and Yann LeCun, editors, {\em 3rd International Conference on Learning Representations, {ICLR} 2015, San Diego, CA, USA, May 7-9, 2015, Conference Track Proceedings}, 2015.

\bibitem{sotgiu2020deep}
Angelo Sotgiu, Ambra Demontis, Marco Melis, Battista Biggio, Giorgio Fumera, Xiaoyi Feng, and Fabio Roli.
\newblock Deep neural rejection against adversarial examples.
\newblock {\em EURASIP Journal on Information Security}, 2020:1--10, 2020.

\bibitem{sperl2020dla}
Philip Sperl, Ching-Yu Kao, Peng Chen, Xiao Lei, and Konstantin B{\"o}ttinger.
\newblock Dla: dense-layer-analysis for adversarial example detection.
\newblock In {\em 2020 IEEE European Symposium on Security and Privacy (EuroS\&P)}, pages 198--215. IEEE, 2020.

\bibitem{springenberg2014striving}
Jost~Tobias Springenberg, Alexey Dosovitskiy, Thomas Brox, and Martin Riedmiller.
\newblock Striving for simplicity: The all convolutional net.
\newblock {\em arXiv preprint arXiv:1412.6806}, 2014.

\bibitem{sturmfels2020visualizing}
Pascal Sturmfels, Scott Lundberg, and Su-In Lee.
\newblock Visualizing the impact of feature attribution baselines.
\newblock {\em Distill}, 5(1):e22, 2020.

\bibitem{subramanya2018towards}
Akshayvarun Subramanya, Vipin Pillai, and Hamed Pirsiavash.
\newblock Towards hiding adversarial examples from network interpretation.
\newblock {\em arXiv preprint arXiv:1812.02843}, 2018.

\bibitem{sundararajan2017axiomatic}
Mukund Sundararajan, Ankur Taly, and Qiqi Yan.
\newblock Axiomatic attribution for deep networks.
\newblock In {\em International conference on machine learning}, pages 3319--3328. PMLR, 2017.

\bibitem{szegedy2013intriguing}
Christian Szegedy, Wojciech Zaremba, Ilya Sutskever, Joan Bruna, Dumitru Erhan, Ian~J. Goodfellow, and Rob Fergus.
\newblock Intriguing properties of neural networks.
\newblock In Yoshua Bengio and Yann LeCun, editors, {\em 2nd International Conference on Learning Representations, {ICLR} 2014, Banff, AB, Canada, April 14-16, 2014, Conference Track Proceedings}, 2014.

\bibitem{tanay2016boundary}
Thomas Tanay and Lewis~D. Griffin.
\newblock A boundary tilting persepective on the phenomenon of adversarial examples.
\newblock {\em CoRR}, abs/1608.07690, 2016.

\bibitem{tao2018attacks}
Guanhong Tao, Shiqing Ma, Yingqi Liu, and Xiangyu Zhang.
\newblock Attacks meet interpretability: Attribute-steered detection of adversarial samples.
\newblock {\em Advances in Neural Information Processing Systems}, 31, 2018.

\bibitem{tramer2022detecting}
Florian Tram{\`{e}}r.
\newblock Detecting adversarial examples is (nearly) as hard as classifying them.
\newblock In {\em International Conference on Machine Learning, {ICML} 2022, 17-23 July 2022, Baltimore, Maryland, {USA}}, volume 162 of {\em Proceedings of Machine Learning Research}, pages 21692--21702. {PMLR}, 2022.

\bibitem{tramer2020adaptive}
Florian Tramer, Nicholas Carlini, Wieland Brendel, and Aleksander Madry.
\newblock On adaptive attacks to adversarial example defenses.
\newblock {\em Advances in neural information processing systems}, 33:1633--1645, 2020.

\bibitem{van9visualizing}
LJP van~der Maaten and GE~Hinton.
\newblock Visualizing high-dimensional data using t-sne. 2008.
\newblock {\em Journal of Machine Learning Research}, 9:2579.

\bibitem{vardhan2021exad}
Raj Vardhan, Ninghao Liu, Phakpoom Chinprutthiwong, Weijie Fu, Zhenyu Hu, Xia~Ben Hu, and Guofei Gu.
\newblock Exad: An ensemble approach for explanation-based adversarial detection.
\newblock {\em arXiv preprint arXiv:2103.11526}, 2021.

\bibitem{wang2020interpretability}
Jingyuan Wang, Yufan Wu, Mingxuan Li, Xin Lin, Junjie Wu, and Chao Li.
\newblock Interpretability is a kind of safety: An interpreter-based ensemble for adversary defense.
\newblock In {\em Proceedings of the 26th ACM SIGKDD International Conference on Knowledge Discovery \& Data Mining}, pages 15--24, 2020.

\bibitem{wu2021beating}
Yuhang Wu, Sunpreet~S Arora, Yanhong Wu, and Hao Yang.
\newblock Beating attackers at their own games: Adversarial example detection using adversarial gradient directions.
\newblock In {\em Proceedings of the AAAI Conference on Artificial Intelligence}, volume~35, pages 2969--2977, 2021.

\bibitem{xie2019feature}
Cihang Xie, Yuxin Wu, Laurens van~der Maaten, Alan~L Yuille, and Kaiming He.
\newblock Feature denoising for improving adversarial robustness.
\newblock In {\em Proceedings of the IEEE/CVF conference on computer vision and pattern recognition}, pages 501--509, 2019.

\bibitem{xu2017feature}
Weilin Xu, David Evans, and Yanjun Qi.
\newblock Feature squeezing: Detecting adversarial examples in deep neural networks.
\newblock In {\em 25th Annual Network and Distributed System Security Symposium, {NDSS} 2018, San Diego, California, USA, February 18-21, 2018}. The Internet Society, 2018.

\bibitem{yang2020ml}
Puyudi Yang, Jianbo Chen, Cho-Jui Hsieh, Jane-Ling Wang, and Michael Jordan.
\newblock Ml-loo: Detecting adversarial examples with feature attribution.
\newblock In {\em Proceedings of the AAAI Conference on Artificial Intelligence}, volume~34, pages 6639--6647, 2020.

\bibitem{zhang2018detecting}
Chiliang Zhang, Zhimou Yang, and Zuochang Ye.
\newblock Detecting adversarial perturbations with salieny.
\newblock In {\em Proceedings of the 6th International Conference on Information Technology: IoT and Smart City}, pages 25--30, 2018.

\bibitem{zhang2018limitations}
Huan Zhang, Hongge Chen, Zhao Song, Duane Boning, Inderjit~S Dhillon, and Cho-Jui Hsieh.
\newblock The limitations of adversarial training and the blind-spot attack.
\newblock In {\em International Conference on Learning Representations}, 2018.

\bibitem{zhang2020interpretable}
Xinyang Zhang, Ningfei Wang, Hua Shen, Shouling Ji, Xiapu Luo, and Ting Wang.
\newblock Interpretable deep learning under fire.
\newblock In Srdjan Capkun and Franziska Roesner, editors, {\em 29th {USENIX} Security Symposium, {USENIX} Security 2020, August 12-14, 2020}, pages 1659--1676. {USENIX} Association, 2020.

\end{thebibliography}

\appendices

\section{Datasets}\label{appendix:dataset}

\textbf{1. MNIST \cite{lecun1998gradient}:} MNIST is a handwritten digit dataset with digits from $0$ to $9$ in 10 classes with 60000 train images and 10000 test images. Each image in the dataset is a grayscale image with a size of $28$x$28$. We use the PyTorch torch-vision MNIST dataset for our evaluation.

\textbf{2. CIFAR-10 \cite{krizhevsky2009learning}:} The CIFAR-10 dataset consists of $32$x$32$ three-channel images in 10 classes with 6000 images per class. There are 50,000 training images and 10,000 test images. The images belong to different real-life objects like airplanes, cars, birds, cats, dogs, frogs, and trucks. We use the PyTorch torch-vision CIFAR-10 dataset for our evaluation.

\textbf{3. CIFAR-100 \cite{krizhevsky2009learning}:} This dataset is similar to CIFAR-10, but it has 100 classes containing 600 images each. There are 500 training images and 100 testing images per class.  We use the PyTorch torch-vision CIFAR-10 dataset for our evaluation.

\textbf{4. ImageNet \cite{deng2009imagenet}:} ImageNet consists of $1000$ classes of high-dimensional real-life RGB images. We use the open-source ImageNet subset available on Kaggle~\cite{kaggle}. It consists of 25000 images on the train set and 3000 images on the validation set.

\textbf{5. Updated CIC-IDS2017 dataset \cite{engelen2021troubleshooting}:} The CIC-IDS2017 dataset \cite{sharafaldin2018toward} is a popular dataset for evaluating intrusion detection systems (IDSs) in network security. This dataset was created by the Canadian Institute for Cybersecurity (CIC) and consists of network traffic data collected in a controlled environment. However, a recent study \cite{engelen2021troubleshooting} demonstrated problems with the feature extraction and labeling of this dataset, and provided an improved version of the dataset\footnote{\url{https://intrusion-detection.distrinet-research.be/WTMC2021/Dataset/dataset.zip}}. We utilize this updated version of CIC-IDS2017 to perform analysis on network intrusion detection. Unlike image data, preprocessing is required for the security dataset available in CSV format. The dataset consists of potentially incorrect values and varying formats, which necessitate preprocessing steps. We transform the categorical features into binary features using one-hot encoding and scale the values within the range of [0, 1] using min-max normalization. Since the goal of adversarial attack on network dataset is to classify malicious traffic flow to benign, we transform the class labels to binary classes (0 for normal traffic and 1 for attack traffic).

\section{Target models}\label{appendix:models}

\subsection{LeNet \cite{lecun1998gradient}}
Lenet is one of the earliest convolutional neural network architectures originally designed for handwritten digit recognition. LeNet consists of a series of convolutional and pooling layers followed by a fully connected layer and output layer for classification. We applied LeNet architecture, as demonstrated in Table \ref{tab:architecturemnist}, for MNIST classification. The model was trained with a learning rate of 0.001 for 60 epochs using a batch size of 64 and the Adam optimizer. We obtained an accuracy of 98.17\% on the test set.

\begin{table}[h]
\centering
\caption{MNIST model architecture}
\label{tab:architecturemnist}
\resizebox{0.48\textwidth}{!}{%
\begin{tabular}{@{}lll@{}}
\toprule
\textbf{\#} & \textbf{Layer} & \textbf{Description} \\ \midrule
1 & Conv2D+ReLU & 6 filters, Kernel size = (5,5), Stride = (1,1) \\
2 & MaxPooling & Kernel size = 2, Stride = 2, Padding = 0 \\
3 & Conv2D+ReLU & 16  filters, Kernel size = (5,5), Stride = (1,1) \\
4 & MaxPooling & Kernel size = 2, Stride = 2, Padding = 0 \\
5 & Dense+ReLU & 256 units \\
6 & Dense+ReLU & 120 units \\
7 & Dense+Softmax & 84 units \\ \bottomrule
\end{tabular}%
}
\end{table}

\subsection{VGG \cite{simonyan2014very}}
VGG networks are also convolutional neural networks with deeper stacking of convolutional layers than LeNet. It consists of a series of convolutional neural networks followed by pooling and fully connected layers. Table \ref{tab:vgg16} summarizes the architecture used for CIFAR-10 classification. The model was trained with a learning rate of 0.001 for 100 epochs using a batch size of 64 and the SGD optimizer with momentum of 0.9 and weight decay of 5e-4. We obtained an accuracy of 84.91\% on the test set.

\begin{table}[h]
\centering
\caption{CIFAR-10 VGG16 Architecture}
\label{tab:vgg16}
\resizebox{0.50\textwidth}{!}{%
\begin{tabular}{@{}lll@{}}
\toprule
\textbf{\#} & Layer & \textbf{Description} \\ \midrule
1 & Conv2d+ReLU & 64 filters, Kernel size = (3, 3) size, Stride=(1,1), Padding = (1,1) \\
2 & Conv2d+ReLU & 64 filters, Kernel size = (3, 3) size, Stride=(1,1), Padding = (1,1) \\
3 & MaxPooling & Kernel size = 2, Stride =2, Padding = 0 \\
4 & Conv2d+ReLU & 128 filters, Kernel size = (3, 3) size, Stride=(1,1), Padding = (1,1) \\
5 & Conv2d+ReLU & 128 filters, Kernel size = (3, 3) size, Stride=(1,1), Padding = (1,1) \\
6 & MaxPooling & Kernel size = 2, Stride =2, Padding = 0 \\
7 & Conv2d+ReLU & 256 filters, Kernel size = (3, 3) size, Stride=(1,1), Padding = (1,1) \\
8 & Conv2d+ReLU & 256 filters, Kernel size = (3, 3) size, Stride=(1,1), Padding = (1,1) \\ 
9 & Conv2d+ReLU & 256 filters, Kernel size = (3, 3) size, Stride=(1,1), Padding = (1,1) \\
10 & MaxPooling & Kernel size = 2, Stride =2, Padding = 0 \\
11 & Conv2d+ReLU & 512 filters, Kernel size = (3, 3) size, Stride=(1,1), Padding = (1,1) \\
12 & Conv2d+ReLU & 512 filters, Kernel size = (3, 3) size, Stride=(1,1), Padding = (1,1) \\
13 & Conv2d+ReLU & 512 filters, Kernel size = (3, 3) size, Stride=(1,1), Padding = (1,1) \\
14 & MaxPooling & Kernel size = 2, Stride =2, Padding = 0 \\
15 & Conv2d+ReLU & 512 filters, Kernel size = (3, 3) size, Stride=(1,1), Padding = (1,1) \\
16 & Conv2d+ReLU & 512 filters, Kernel size = (3, 3) size, Stride=(1,1), Padding = (1,1) \\
17 & Conv2d+ReLU & 512 filters, Kernel size = (3, 3) size, Stride=(1,1), Padding = (1,1) \\
18 & MaxPooling & Kernel size = 2, Stride =2, Padding = 0 \\
19 & Average Pooling & Kerne size = 1, Stride = 1, Padding = 0 \\
20 & Dense+Softmax & 512 units \\ \bottomrule
\end{tabular}%
}
\end{table}

\subsection{ResNet \cite{he2016deep}}
ResNet, short for ``Residual Networks," is a deep convolutional neural network. The distinguishing characteristic of ResNet is the use of residual blocks. A residual block consists of a ``shortcut" or ``skip connection" that bypasses one or more convolutional layers. This shortcut allows the network to learn residual functions, the difference between the desired output and the actual output of the layer. This skip connection enables the training of extremely deep networks without the vanishing gradient problem.  We used ResNet for CIFAR-10, CIFAR-100, and ImageNet datasets. Depending on the depth of the network, ResNet is further represented in variants like ResNet-18, ResNet-50, ResNet-56, ResNet-152.  For CIFAR-100, we used a pre-trained ResNet56 model from PyTorch \cite{he2016deep}. It achieved 64.43\% accuracy on the test set. For CIFAR-10, we trained ResNet18 \cite{he2016deep} models, which achieved 92.5\% accuracy on the test set. For ImageNet, we used a pre-trained Resnet50 \cite{he2016deep} model from the Torch library, which achieved 76.13\% accuracy on the test set.

\subsection{MobileNet \cite{sandler2018mobilenetv2}}
MobileNet is a family of network architectures designed for efficient deep learning on mobile and embedded devices by minimizing computational and memory resources. We use the MobileNetV2 model available in PyTorch. MobileNetV2 introduces "inverted residuals" layers, which consist of bottleneck layers, shortcut connections, and linear bottlenecks. Each inverted residual block includes an expansion layer, which increases the number of channels before the depth-wise separable convolution. MobileNet relies on depth-wise separable convolution, which reduces the computational cost by separating the convolution process into depth-wise and point-wise convolutions. For ImageNet, we used a pre-trained MobileNet \cite{sandler2018mobilenetv2} from Torch library, which achieved 70.1\% accuracy on the test set.

We summarize the dataset, architecture, and their performance on the test in Table \ref{tab:datasetarchitecture}. 
\begin{table}[]
\centering
\caption{Datasets and DNN Architectures.}
\label{tab:datasetarchitecture}
\resizebox{0.48\textwidth}{!}{%
\begin{tabular}{@{}cccc@{}}
\toprule
\multicolumn{1}{c}{\textbf{Dataset}} & \textbf{Number of classes} & \textbf{Test Accuracy} & \textbf{Architecture} \\ \midrule
MNIST  & 10 & 98.17\% & LeNet \cite{lecun1998gradient} \\
CIFAR-10  & 10 & 92.5\% & ResNet \cite{he2016deep} \\
CIFAR-10 & 10 & 84.91\% & VGG \cite{simonyan2014very} \\ 
CIFAR-100 & 100 & 64.43\% & ResNet \cite{he2016deep}\\ 
ImageNet  & 1000 & 76.13\% & ResNet \cite{he2016deep} \\
ImageNet  & 1000 &  70.1\% & MobileNet \cite{sandler2018mobilenetv2} \\
CIC-IDS2017  & 2 &  80.18\% & MLP \cite{rumelhart1985llearning} \\ \bottomrule
\end{tabular}}
\end{table}

\subsection{Network Intrusion Detection}\label{apendix:networkintrusion}

Similar to prior works \cite{engelen2021troubleshooting}, we use a Multi-Layer Perceptron (MLP) network as an intrusion detector. It consists of 2 hidden layers of 64 neurons and a softmax output layer with 2 neurons. Each neuron in the hidden layer uses ReLU activation. We train the model for 1000 epochs using the Adam optimizer, and the learning rate of 0.01. 

\section{Adversarial attack}\label{appendix:attack}
\textbf{Fast Gradient Sign Attack (FGSM) \cite{goodfellow2015explaining}:} FGSM is a computationally efficient method for finding adversarial examples. It assumes a linear approximation of the network loss function and finds a perturbation by increasing the local linear approximation of the loss. The perturbation for an FGSM attack against a network with loss $J$ and parameters $\theta$, for test sample $\textbf{x}$, and with true label $y$ is given by:
 \begin{equation}\label{eqn:fgsm}
     \delta=\epsilon * \textnormal{sign}(\nabla_{x}J(\theta, \textbf{x}, y))
 \end{equation}
 
 The strength of the perturbation for each dimension of the input is controlled by $\epsilon$. We use $\epsilon \in [8/255,16/255,32/255,64/255]$.

\textbf{Basic Iterative Method (BIM) \cite{kurakin2018adversarial}}: BIM is an iterative version of FGSM where the perturbation is computed multiple times with small steps. The pixel values of the resulting adversarial image are clipped to ensure they lie within the $L_\infty$ $\epsilon$ neighborhood of the original input image. 
\begin{equation}
    \textbf{x}^{*}_{m+1} = \textbf{x}^{*}_{m} + Clip_{\textbf{x}, \epsilon} (\alpha . \textnormal{sign}(\nabla_{x}J(\theta, \textbf{x}^{*}_{m}, y)) 
\end{equation}

Here, ${0<\alpha<\epsilon}$ controls the $m^{th}$ iteration step size. We use $\epsilon \in [8/255,16/255,32/255,64/255]$ with $\alpha = \epsilon/10$ and a fixed number of iterations ($m$) as 10.  

\textbf{Projected Gradient Descent (PGD) \cite{madry2017towards}:} PGD is also an iterative method similar to BIM; however, unlike BIM, a PGD attack starts from a random perturbation in the $L_\infty$ ball around the input sample.
 \begin{equation} \label{eqn:iterativegradient}
     \textbf{x}^*=\textbf{x}_{n-1}-\textnormal{clip}_{\epsilon}(\alpha~\textnormal{sign}(\nabla xJ(\theta,\textbf{x}_{n-1},y))
 \end{equation}

 We use $\epsilon \in [8/255,16/255,32/255,64/255]$ with $\alpha = \epsilon/10$ and attack steps as $c*\frac{\epsilon}{\alpha}$ where $c$ is a constant. We use $c=4$, so we apply an attack step of 40 for PGD attack across different $\epsilon$.

\textbf{Auto Projected Gradient Descent (Auto-PGD) \cite{croce2020reliable}:}  Auto-PGD attack is gradient-based adversarial attack that builds upon the Projected Gradient Descent (PGD) metohd. It aims to automate the process of finding effective attack parameters, reducing the need for manual tuning of step size and other hyperparameters. Auto-PGD uses an alternative loss function for better performance against defenses that might attempt to mask gradients. We use the Auto-PGD implementation available in Adversarial Robustness Toolbox (ART) \cite{art2018}. An $\epsilon= 0.15$ is considered a relatively moderate to strong attack strength, which we choose in this paper.

\textbf{Carlini and Wagner (CW) \cite{carlini2017towards}:} CW attacks comprise a range of attacks that follow an optimization framework similar to L-BFGS \cite{szegedy2013intriguing}. However, it replaces the loss function with an optimization problem involving logits $(Z(.))$ instead of the model prediction.
\begin{equation}
    g(\textbf{x}) = max(max_({i\neq t} Z(\textbf{x}')_i)-Z(\textbf{x}')_t, -k)
\end{equation}

Here, $k$ encourages the optimizer to find an adversarial example with high confidence. For $L_\infty$ CW attack, we use $\epsilon=0.15$, 400 iterations, and zero confidence settings. We use a learning rate of 0.01 for the step size of the optimization.

\section{Implementation of various detectors}\label{appendix:implementation}

\textbf{Feature squeezing (FS) \cite{xu2017feature}:} For MNIST, we use bit depth reduction and median filter, while for CIFAR-10, CIFAR-100, and ImageNet, we use bit depth reduction, median filter, and non-local means, as suggested by the original paper.

\textbf{Magnet \cite{meng2017magnet}:} We use the defensive architecture recommended by the original paper for CIFAR-10 and MNIST. Since the original paper did not perform an evaluation on CIFAR-100 and ImageNet, we use the defensive architecture of CIFAR-10, which is designed for three-channel images.

\textbf{Turning a weakness into a strength (TWS) \cite{hu2019new}:}  We follow the implementation shared by the authors available on Github\footnote{\url{https://github.com/s-huu/TurningWeaknessIntoStrength}}. We use Gaussian noise of $\sigma = 0.01$ and $\sigma = 0.1$ for CIFAR and ImageNet, as suggested in the paper. For MNIST and CIFAR-100, we empirically picked noise parameters ($\sigma = 0.01$ and $\sigma = 0.1$) that resulted in maximum adversarial detection.

\textbf{U-LOO \cite{yang2020ml}:} For each data set, we randomly select 2000 benign samples, extract feature attribution using the Integrated Gradient method, and compute the inter-quartile range (IQR). IQR is the difference between the 75th percentile and the 25th percentile among all entries of $IG(x) \in R^d$. A sample is regarded as adversarial if the IQR is larger than the threshold learned from benign samples.

\section{Indicator of Failure: PGD at 100 steps}\label{appendix:pgd100steps}
Following Pintor et al.\cite{pintor2022indicators} recommendation, we reevaluated detection methods on PGD adversarial samples crafted with 100 iteration steps. We set the attack parameter $\epsilon = 8/255$. We can observe in Table \ref{tab:attackFail} that most detection methods do not have significant changes to the detection performance when compared with prior performance on detecting adversarial samples crafted at steps of 40.

\begin{table}[]
\centering
\caption{Evaluation of PGD attack with steps = 100}
\label{tab:attackFail}
\resizebox{0.46\textwidth}{!}{%
\begin{tabular}{@{}lcccccc@{}}
\toprule
\textbf{}                     & \textbf{AUC}    & \textbf{FS} & \textbf{MagNet} & \textbf{U-LOO} & \textbf{TWS} & \textbf{PASA} \\ \midrule
\textbf{MNIST}             & \textbf{Before} & 0.90$\pm$0.01 & 0.95$\pm$0.03 & 0.99$\pm$0.01 & 0.92$\pm$0.11 & 0.98$\pm$0.01 \\
\textbf{}                  & \textbf{After}  & 0.87$\pm$0.02 & 0.95$\pm$0.03 & 0.99$\pm$0.99 & 0.90$\pm$0.08 & 0.97$\pm$0.02 \\ \midrule
\textbf{CIFAR-10 (VGG)}    & \textbf{Before} & 0.52$\pm$0.01 & 0.59$\pm$0.03 & 0.49$\pm$0.05 & 0.58$\pm$0.02 & 0.74$\pm$0.02 \\
                           & \textbf{After}  & 0.32$\pm$0.02 & 0.58$\pm$0.01 & 0.48$\pm$0.03 & 0.11$\pm$0.02 & 0.75$\pm$0.02 \\ \midrule
\textbf{CIFAR-10 (ResNet)} & \textbf{Before} & 0.25$\pm$0.01 & 0.57$\pm$0.04 & 0.62$\pm$0.01 & 0.14$\pm$0.01 & 0.83$\pm$0.03 \\
\textbf{}                  & \textbf{After}  & 0.24$\pm$0.01 & 0.56$\pm$0.04 & 0.64$\pm$0.01 & 0.14$\pm$0.01 & 0.83$\pm$0.02 \\ \midrule
\textbf{CIFAR-100}         & \textbf{Before} & 0.67$\pm$0.03 & 0.56$\pm$0.03 & 0.63$\pm$0.03 & 0.61$\pm$0.01 & 0.60$\pm$0.02 \\
                           & \textbf{After}  & 0.68$\pm$0.02 & 0.55$\pm$0.02 & 0.60$\pm$0.03 & 0.67$\pm$0.02 & 0.58$\pm$0.03 \\ \midrule
\textbf{ImageNet (MobileNet)} & \textbf{Before} & 0.25$\pm$0.02        & 0.51$\pm$0.01            & 0.59$\pm$0.01           & 0.52$\pm$0.01         & 0.98$\pm$0.02          \\
\textbf{}                  & \textbf{After}  & 0.28$\pm$0.02 & 0.49$\pm$0.02 & 0.51$\pm$0.01 & 0.50$\pm$0.02 & 0.97$\pm$0.01 \\ \midrule
\textbf{ImageNet (ResNet)} & \textbf{Before} & 0.29$\pm$0.01 & 0.51$\pm$0.01 & 0.59$\pm$0.01 & 0.57$\pm$0.02 & 0.97$\pm$0.01 \\
                           & \textbf{After}  & 0.27$\pm$0.02 & 0.49$\pm$0.01 & 0.59$\pm$0.01 & 0.59$\pm$0.02 & 0.96$\pm$0.01 \\ \bottomrule
\end{tabular}%
}
\end{table}

\section{Ablation study}\label{appendix:ablation}

Our detection method combines two statistical metrics: prediction sensitivity (PS) and attribution sensitivity (AS). In this ablation study, we assess the individual performance of each metric. We summarize the results in Table \ref{tab:ablationmnistcifarappendix} and Table \ref{tab:ablationimagenetappendix}. 

In Table \ref{tab:ablationmnistcifarappendix}, we can observe that the performance of each metric is almost equivalent to the combined performance. However, on CIFAR-10, the combination of AS and PS (PS+AS) consistently outperforms AS and PS individually. AS and PS exhibit sensitivity to different attack types. For instance, PS is more effective at detecting adversarial inputs generated by FGSM, while AS excels in detecting inputs perturbed by PGD and other attacks in CIFAR-10. Combining both metrics provides a more balanced and robust detection strategy across various attack types.

\begin{table}[]
\centering
\caption{Adversarial detection performance for MNIST and CIFAR. Here, AS represents Attribution Sensitivity, and PS represents Prediction Sensitivity. PS+AS is our proposed detector.}
\label{tab:ablationmnistcifarappendix}
\resizebox{0.5\textwidth}{!}{%
\begin{tabular}{@{}llccc|ccc|ccc@{}}
\toprule
\multicolumn{1}{c}{} &
  \multicolumn{1}{c}{} &
  \multicolumn{3}{c|}{\textbf{MNIST}} &
  \multicolumn{3}{c|}{\textbf{CIFAR-10 (ResNet)}} &
  \multicolumn{3}{c}{\textbf{CIFAR-10 (VGG)}} \\ \midrule
\textbf{Type} &
  \multicolumn{1}{c}{\textbf{Strength}} &
  \textbf{PS} &
  \textbf{AS} &
  \textbf{PS+AS} &
  \textbf{PS} &
  \textbf{AS} &
  \textbf{PS+AS} &
  \textbf{PS} &
  \textbf{AS} &
  \textbf{PS+AS} \\ \midrule
\textbf{FGSM} & \textbf{8/255}  & 0.97 & 0.84 & 0.97 & 0.77 & 0.87 & 0.88 & 0.54 & 0.55 & 0.62 \\
              & \textbf{16/255} & 0.98 & 0.85 & 0.98 & 0.96 & 0.97 & 0.99 & 0.67 & 0.51 & 0.71 \\
              & \textbf{32/255} & 0.98 & 0.86 & 0.98 & 0.99 & 0.99 & 0.99 & 0.87 & 0.69 & 0.87 \\
              & \textbf{64/255} & 0.99 & 0.85 & 0.98 & 0.98 & 0.99 & 0.99 & 0.94 & 0.89 & 0.94 \\ \midrule
\textbf{PGD}  & \textbf{8/255}  & 0.98 & 0.85 & 0.98 & 0.13 & 0.88 & 0.85 & 0.51 & 0.57 & 0.61 \\
              & \textbf{16/255} & 0.98 & 0.87 & 0.97 & 0.30  & 0.94 & 0.92 & 0.58 & 0.51 & 0.61 \\
              & \textbf{32/255} & 0.98 & 0.89 & 0.98 & 0.22 & 0.97 & 0.97 & 0.74 & 0.51 & 0.73 \\
              & \textbf{64/255} & 0.98 & 0.89 & 0.98 & 0.08 & 0.98 & 0.98 & 0.89 & 0.81 & 0.90  \\ \midrule
\textbf{BIM}  & \textbf{8/255}  & 0.56 & 0.46 & 0.58 & 0.09 & 0.83 & 0.77 & 0.49 & 0.54 & 0.58 \\
              & \textbf{16/255} & 0.55 & 0.47 & 0.52 & 0.09 & 0.89 & 0.84 & 0.52 & 0.53 & 0.59 \\
              & \textbf{32/255} & 0.51 & 0.42 & 0.53 & 0.16 & 0.94 & 0.93 & 0.55 & 0.47 & 0.60  \\
              & \textbf{64/255} & 0.54 & 0.39 & 0.54 & 0.16 & 0.97 & 0.98 & 0.70  & 0.49 & 0.68 \\ \midrule
\textbf{CW}   & \textbf{0.15}   & 0.38 & 0.34 & 0.38 & 0.95 & 0.97 & 0.98 & 0.79 & 0.61 & 0.80  \\ \bottomrule
\end{tabular}%
}
\end{table}

In Table \ref{tab:ablationimagenetappendix}, both attribution sensitivity and prediction sensitivity have high detection performance in detecting FGSM attacks. However, with PGD, BIM, and CW, individual metrics have weaker performances. The combination of AS and PS (PS+AS) consistently outperforms the individual metrics (AS and PS). The detector's performance generally degrades as the strength of adversarial attacks increases. This degradation is more pronounced in cases where the AS and PS metrics are employed individually, noticeable with ImageNet.

\begin{table}[]
\centering
\caption{Adversarial detection performance for CIFAR-100 and ImageNet. Here, AS represents Attribution Sensitivity, and PS represents Prediction Sensitivity. PS+AS is our proposed detector.}
\label{tab:ablationimagenetappendix}
\resizebox{0.5\textwidth}{!}{%
\begin{tabular}{@{}llccc|ccc|ccc@{}}
\toprule
 &
   &
  \multicolumn{3}{l|}{\textbf{CIFAR 100}} &
  \multicolumn{3}{l|}{\textbf{ImageNet (Mobilenet)}} &
  \multicolumn{3}{l}{\textbf{ImageNet (ResNet)}} \\ \midrule
\textbf{Type} &
  \multicolumn{1}{c}{\textbf{Strength}} &
  \textbf{PS} &
  \textbf{AS} &
  \textbf{PS+AS} &
  \textbf{PS} &
  \textbf{AS} &
  \textbf{PS+AS} &
  \textbf{PS} &
  \textbf{AS} &
  \textbf{PS+AS} \\ \midrule
\textbf{FGSM} & \textbf{8/255}  & 0.68 & 0.74 & 0.82 & 0.62 & 0.81 & 0.82 & 0.48 & 0.60  & 0.65 \\
              & \textbf{16/255} & 0.78 & 0.89 & 0.95 & 0.68 & 0.91 & 0.91 & 0.55 & 0.73 & 0.75 \\
              & \textbf{32/255} & 0.89 & 0.89 & 0.97 & 0.74 & 0.96 & 0.96 & 0.61 & 0.87 & 0.87 \\
              & \textbf{64/255} & 0.95 & 0.89 & 0.97 & 0.80  & 0.98 & 0.98 & 0.70  & 0.93 & 0.95 \\ \midrule
\textbf{PGD}  & \textbf{8/255}  & 0.62 & 0.38 & 0.57 & 0.82 & 0.35 & 0.97 & 0.78 & 0.23 & 0.97 \\
              & \textbf{16/255} & 0.63 & 0.54 & 0.66 & 0.82 & 0.33 & 0.98 & 0.81 & 0.21 & 0.97 \\
              & \textbf{32/255} & 0.63 & 0.74 & 0.80  & 0.83 & 0.41 & 0.98 & 0.83 & 0.13 & 0.98 \\
              & \textbf{64/255} & 0.62 & 0.88 & 0.94 & 0.82 & 0.62 & 0.98 & 0.82 & 0.19 & 0.97 \\ \midrule
\textbf{BIM}  & \textbf{8/255}  & 0.56 & 0.36 & 0.57 & 0.41 & 0.26 & 0.49 & 0.25 & 0.27 & 0.35 \\
              & \textbf{16/255} & 0.59 & 0.40  & 0.61 & 0.51 & 0.23 & 0.64 & 0.34 & 0.19 & 0.46 \\
              & \textbf{32/255} & 0.61 & 0.59 & 0.69 & 0.64 & 0.21 & 0.76 & 0.46 & 0.18 & 0.59 \\
              & \textbf{64/255} & 0.57 & 0.79 & 0.83 & 0.69 & 0.28 & 0.84 & 0.47 & 0.16 & 0.63 \\ \midrule
\textbf{CW}   & \textbf{0.15}   & 0.62 & 0.85 & 0.92 & 0.81 & 0.44 & 0.87 & 0.77 & 0.37 & 0.94 \\ \bottomrule
\end{tabular}%
}
\end{table}

\end{document}